\documentclass[12pt,preprint]{aastex}
\usepackage{latexsym}% 
\usepackage{graphicx}% 
\usepackage{amssymb}% 
\usepackage{longtable}%
\def\kms{\ifmmode{\rm km\thinspace s^{-1}}\else km\thinspace s$^{-1}$\fi}

\begin{document}

\title{The TWA 3 Young Triple System: Orbits, Disks, Evolution}

\author{Kendra Kellogg\altaffilmark{1}, L. Prato\altaffilmark{2}, Guillermo Torres\altaffilmark{3}, G. H. Schaefer\altaffilmark{4}, 
I. Avilez\altaffilmark{2}, D. Ru\'{i}z-Rodr\'{i}guez\altaffilmark{5}, L. H. Wasserman\altaffilmark{2}, Alceste Z. Bonanos\altaffilmark{6}, E. W. Guenther\altaffilmark{7},
R. Neuh\"auser\altaffilmark{8}, S. E. Levine\altaffilmark{2,9}, A. S. Bosh\altaffilmark{9,2}, Katie M. Morzinski\altaffilmark{10}, Laird Close\altaffilmark{10},
Vanessa Bailey\altaffilmark{11}, Phil Hinz\altaffilmark{10}, Jared R. Males\altaffilmark{10}}

\altaffiltext{1}{Department of Physics and Astronomy, The University of Western Ontario,
London, ON N6A 3K7, Canada; kkellogg@uwo.ca}
\altaffiltext{2}{Lowell Observatory, 1400 West Mars Hill Road, Flagstaff,
AZ 86001, USA; lprato@lowell.edu}
\altaffiltext{3}{Harvard-Smithsonian Center for Astrophysics, Cambridge, MA 02138, USA}
\altaffiltext{4}{The CHARA Array of Georgia State University, Mount Wilson Observatory, Mount Wilson, CA 91023, USA}
\altaffiltext{5}{Research School of Astronomy and Astrophysics, Australian National University, Canberra, ACT 2611, Australia}
\altaffiltext{6}{IAASARS, National Observatory of Athens, 15236 Penteli, Greece}
\altaffiltext{7}{Th\"uringer Landessternwarte Tautenburg, 07778 Tautenburg, Germany}
\altaffiltext{8}{Astrophysikalisches Institut und Universit\"ats-Sternwarte, FSU Jena,
Schillerg\"a\ss{}chen 2-3, D-07745 Jena, Germany}
\altaffiltext{9}{MIT Department of Earth, Atmospheric, and Planetary Sciences, Cambridge, MA 02139, USA}
\altaffiltext{10}{Steward Observatory, University of Arizona, 933 N. Cherry Ave, Tucson, AZ 85721, USA}
\altaffiltext{11}{Kavli Institute for Particle Astrophysics and Cosmology, Department of Physics, Stanford University, Stanford, CA, 94305, USA}

\begin{abstract}
We have characterized the spectroscopic orbit of the TWA 3A binary and provide preliminary families of probable solutions
for the TWA 3A visual orbit as well as for the wide TWA 3A--B orbit.  TWA 3 is a hierarchical triple located at 34 pc in the
$\sim$10 Myr old TW Hya association.  The wide component separation is 1\farcs55; the close pair was first identified as
a possible binary almost 20 years ago.  We initially identified the 35-day period orbital solution using high-resolution infrared spectroscopy
which angularly resolved the A and B components.
We then refined the preliminary orbit by combining the infrared data with a re-analysis of our high-resolution optical spectroscopy.
The orbital period from the combined spectroscopic solution is $\sim$35 days, the eccentricity is $\sim$0.63, and the mass ratio is $\sim$0.84; although this
high mass ratio would suggest that optical spectroscopy alone should be sufficient to identify the orbital solution, the presence of the tertiary B
component likely introduced confusion in the blended optical spectra.  Using millimeter imaging from the literature, we also estimate
the inclinations of the stellar orbital planes with respect to the TWA 3A circumbinary disk inclination and find that all three
planes are likely misaligned by at least $\sim$30 degrees.  The TWA 3A spectroscopic binary components have spectral types of
M4.0 and M4.5; TWA 3B is an M3.  We speculate that the system formed as a triple, is bound, and that its properties were shaped by 
dynamical interactions between the inclined orbits and disk.

\end{abstract}

\section{Introduction}
A key question in astronomy involves the mechanisms and characteristic ages 
for the formation of planetary systems like our own. One approach to address this is the study of the evolution
and dynamics of protoplanetary disks -- the sites of planet formation around young stars. Although our own
solar system is centered around a single star, it is nevertheless important
to understand planet formation in multiple star systems, not only because most stars form in
binary or higher order multiples which therefore dominate the star formation process (e.g., Duch{\^e}ne \& Kraus 2013), but also
because a number of exoplanetary systems have been identified in binary
star systems (e.g., Doyle et al. 2011; Kostov et al. 2014; Welsh et al. 2015). Furthermore, young multiple systems
provide a means to examine how tidal interactions affect
circumstellar disks. The TW Hydrae association provides an advantageous observing ground for young
circumstellar disk evolution because at a distance of only $<\sim$50 pc, it is the nearest
known group of young stars.  Serendipitously, for a small stellar population
(e.g., Torres et al. 2003) it contains a disproportionately large fraction of 
young multiples, some with complex circumstellar and circumbinary disk configurations
(Webb et al. 1999; Koerner et al. 2000; Prato et al. 2001;
Konopacky et al. 2007; Andrews et al. 2010).

Previous work has shown that whether a disk is circumstellar
or circumbinary, the protoplanetary material usually dissipates
within $\sim$10 Myr and rarely is found to be accreting beyond that age (e.g., Haisch et al. 2001; Hernandez et al. 2008).
TWA 3, however, is one of the systems in the 10 Myr old TW Hydrae association
that is an exception (Muzerolle et al. 2000).  TWA 3 is a visual binary with a 2015 position angle of 207$^{\circ}$
and a projected separation of 1\farcs55 (Tokovinin et al. 2015), corresponding to $\sim$50 AU at a distance of 34 $\pm$4 pc (Mamajek 2005).  Although 
different distance estimates are available in the literature, we have chosen to  adopt Mamajek's estimate given the
consistency with Gagne et al. (2017).  The TW Hya association members demonstrate a wide range of distances, from $\sim$30--50~pc,
consistent with the elongated shape of the region as discussed in Kellogg et al. (2016).
Basic properties of the system are listed in Table 1.
Jayawardhana et al. (1999) angularly resolved the visual pair in the near- and mid-infrared (IR),
demonstrating that TWA 3A exhibits optically thick excess long-wavelength emission,
indicative of a dusty disk.  TWA 3B appears to be devoid of dust.  Furthermore,
TWA 3A also has a significant gas accretion signature similar to TW Hya and TWA 14 (Muzerolle et al. 2000).
Webb et al. (1999) measured H$\alpha$ emission equivalent widths of 21.8 and 7.14 \AA\  for TWA 3A
and 3B, respectively.  Herczeg et al. (2009) found equivalent widths of 37 and 3.4 \AA\  for TWA 3A
and 3B, respectively. Given the 10 \AA\ and 20 \AA\  approximate limits for early and late M stars with active disks
(Mart\'{i}n et al. 1998), the B component is likely not accreting while
the A component is a bona fide, classical T Tauri star (CTTS).  However, Muzerolle et al. (2000)
calculated a relatively low gas accretion rate of $\sim5 \times 10^{-11}$ M$_{\odot}$ yr$^{-1}$ for TWA 3A.
Although Herczeg et al. (2009) used the Balmer continuum emission from TWA 3A to measure a larger
accretion rate of $2.5 \times 10^{-10}$ M$_{\odot}$ yr$^{-1}$, 
typical accretion rates for classical T Tauri stars range from 10$^{-9}$ to 10$^{-7}$~M$_{\odot}$ yr$^{-1}$ 
(Bouvier et al. 2007).  This modest accretion rate and the relatively low-intensity X-ray emission (Huenemoerder et al. 2007),
coupled with indications for a dust gap in the inner disk
(Jayawardhana et al. 1999; Andrews et al 2010), suggest that the disk around the A-component is likely a
pre-transition disk, about to evolve from a CTTS to a weak-lined T Tauri (WTTS) or to a diskless
system like TWA 3B.

Measurements at 10$\mu$m have shown the presence of silicates in the TWA 3A
disk suggesting some grain growth (Uchida et al. 2004).  Andrews et al. (2010) resolved the disk
in continuum measurements at 880 $\mu$m.  Their data show no evidence for warm dust around
the B component or surrounding the entire system.  Combining their data with a broadband spectral energy distribution (SED)
and comparing to models of truncated disks, Andrews et al. estimated
an outer disk radius of $\sim$15--25 AU and an inner radius of $\sim$1 AU,
consistent with the 1.3 AU estimate of Uchida et al. (2004).

Webb et al. (1999) suggested, and Muzerolle et al. (2000) and Torres et al. (2003) confirmed, that TWA 3A is a
spectroscopic binary, but one for which an orbital solution has been elusive.  Substantial
optical data of the unresolved system, described in Torres et al., was collected at
several facilities on this system over 2 decades by several of us.
Following the speculation that A is itself a spectroscopic binary in Webb et al., beginning in 2002
one of us also obtained occasional
observations of the system with the NIRSPEC high-resolution IR spectrograph at
the Keck II telescope.  This approach provided important advantages:  the superior seeing
typical on Mauna Kea, in conjunction with a 10-meter telescope, allowed for straightforward
angular resolution of the visual binary pair on a regular basis.
Furthermore, because luminosity is a steeper
function of mass in the optical compared to the IR, longer wavelength observations
facilitate identification of the spectrum of a fainter and redder secondary
star in the blended lines of the TWA 3A spectroscopic binary (Prato et al. 2002a).

This approach has yielded an orbital solution for the TWA 3A system and a mass ratio
measurement for the pair, increasing the small pool of double-lined solutions for
pre-main sequence spectroscopic binaries (e.g., Rosero et al. 2011).  With the advent of
ALMA and GAIA, increasingly sensitive techniques will provide the potential
means for measuring the total system mass using the velocity curve of a circumbinary
disk (Guilloteau et al. 2014) or using the astrometric motion of a binary
photocenter to measure the orbital inclination (e.g., Goldin \& Makarov 2007).
Angularly resolved visual orbits from adaptive optics (AO) imaging and interferometry
also play an important role (e.g., Simon et al. 2013; Le Bouquin et al. 2014).
In conjunction with the mass ratio, these advances will allow for the determination
of the individual component masses, crucial input to models of young star evolution (Prato et al. 2002b).

The confirmation of the TWA 3A spectroscopic binary demonstrates that its long-lived
disk is circumbinary.  A standard paradigm for disk evolution in young hierarchical triple
systems anticipates star--disk tidal interactions that disrupt
circumbinary material while circumstellar material surrounding the single star
remains intact and continues to accrete onto the central star for a longer period of time (White et al. 2002).  
Although one might expect circumbinary disk disruption to be a function of stellar separation, only
a handful of young spectroscopic binaries have circumbinary material and the orbital periods of these pairs
range from a couple of days to almost a year.
TWA 3 is an intriguing example of a system with both a single and a binary and
a long-lived disk only around the binary; this triple provides potentially valuable clues to
understanding important but subtle aspects of disk evolution around low-mass stars.
In \S 2 we describe the visible light and IR observations and data reduction, and in
\S 3 the details of the radial velocity (RV) analysis and results.  We provide a description in \S 4
of the hierarchical visual and spectroscopic orbits in the system.  A description of the structure
and origin of this complex system appears in
\S 5; these results are discussed in \S 6 and \S 7 provides a summary.

\section{Observations}

We report on close to two decades of imaging and spectroscopy of the TWA 3 system.
Table 2 lists the facilities and instrumentation which supplied the data sets described below.

\subsection{Near-Infrared}

\subsubsection{Spectroscopy}

Near-IR, spectroscopic observations were made over eleven epochs between 2002 December and 2011
February with the NIRSPEC instrument on the Keck II 10m telescope, located on Mauna Kea. The UT
dates of observation are listed in Table 3. NIRSPEC is a near-IR, cross-dispersed,
cryogenic spectrograph and employs a $1024 \times 1024$ ALADDIN InSb array detector.
Our data were taken in the $H$-band with a central wavelength of $\sim$1.555 $\mu$m. We used the
0\farcs288 slit, yielding a resolution of $R \approx 30,\!000$. TWA 3 has a Two Micron All Sky Survey (2MASS) $H$-band
magnitude of 7.041 (Table 1).  Integration times varied between 20s and 180s, depending on seeing
conditions and whether or not NIRSPEC was employed behind the AO system, which
reduces the total throughput by a factor of $\sim$5 because of the addition of numerous reflecting surfaces on the AO bench.
We nodded the telescope between two positions on the slit in an AB or
an ABBA nod sequence to allow for background subtraction between sequential spectra. 

\begin{figure}
\centering
\includegraphics[width=6in]{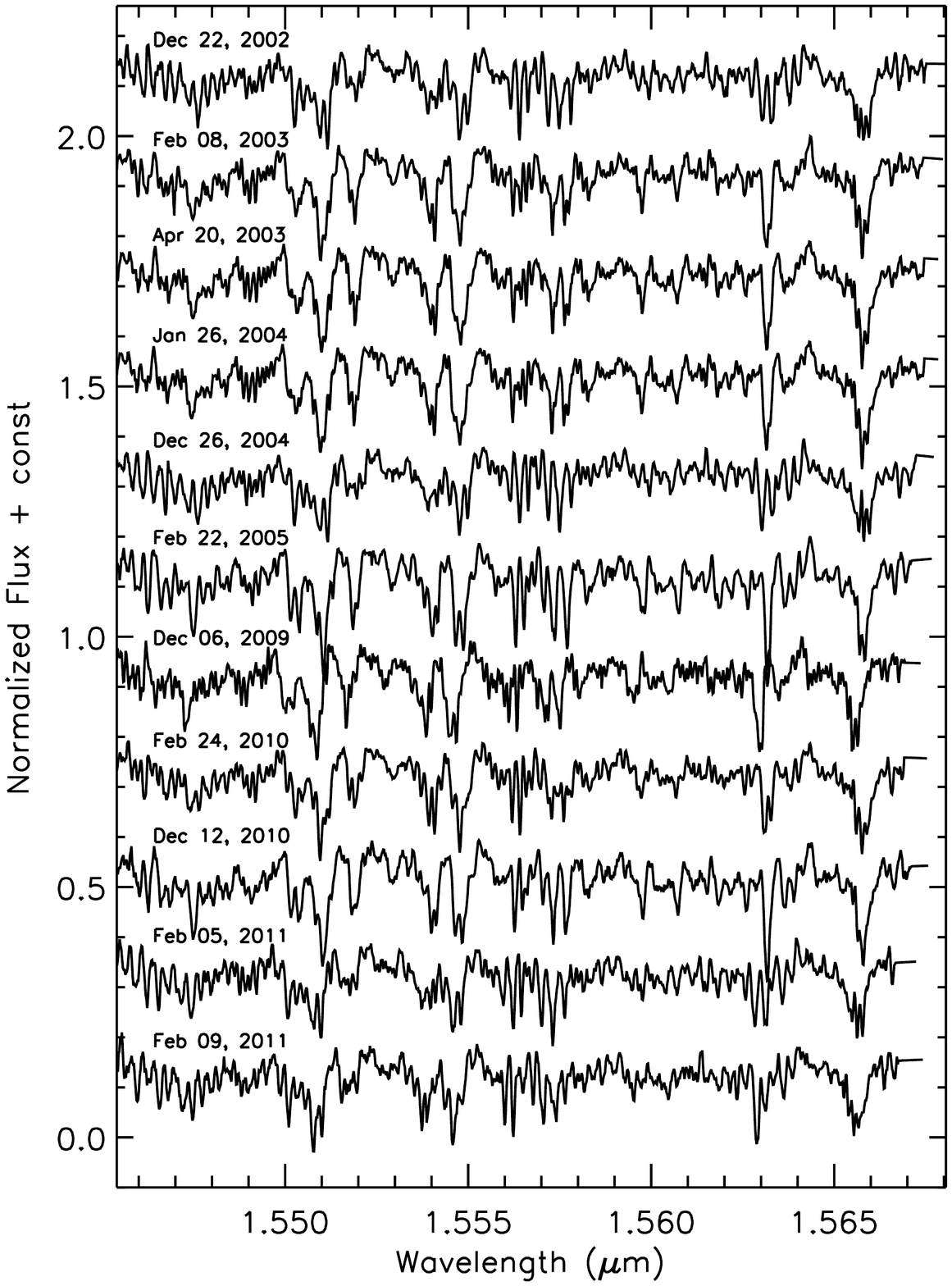}
\caption{Near-IR spectra of TWA 3A from all 11 epochs.
The spectra have been normalized to unity, barycentric corrected,
and displaced along the vertical axis for viewing purposes by
the addition of a constant.  The UT dates of observation are given
above each spectrum.  Effectively every detectable feature is real;
the signal-to-noise ratio ranges from $\sim$150 to several hundred.
The doubling of the lines is evident in at least half of the epochs.}
\label{asa}
\end{figure}

\begin{figure}
\centering
\includegraphics[width=6in]{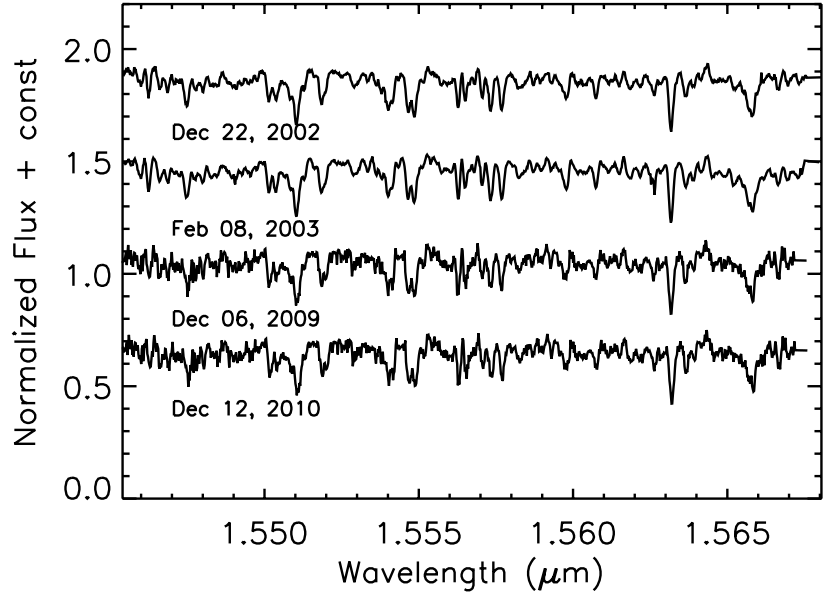}
\caption{Spectra of TWA 3B from all epochs, plotted as the 3A spectra
in Figure 1.  The signal-to-noise in the 2009 and 2010 data is $\sim$100
yet again all the discernible features are likely real.}
\label{asb}
\end{figure}

All near-IR spectroscopic data reductions were made with the REDSPEC
package\footnote{See: http://www2.keck.hawaii.edu/inst/nirspec/redspec/index.html} (Kim, Prato, \& McLean 2015).
We analyzed an order centered around 1.555 $\mu$m in the $H$-band, a convenient region
because it is devoid of all atmospheric absorption lines, therefore eliminating
the need to divide by a telluric standard star in the reduction process. There are also
strong OH night sky emission lines spaced at relatively regular intervals across this
order (Rousselot et al. 2000), useful for determining the spectral dispersion solution. The
procedure for using REDSPEC is described in detail by Prato et al. (2002a).  In brief,
a median-filtered cube of dark frames was subtracted from a median-filtered cube of flat
frames to create a master flat which was then divided into A--B subtracted pairs.
Using fits to the spectral traces and the night sky OH emission lines or comparison lamp
lines, the software rectified this semi-processed frame and determined the
wavelength zero-point and dispersion solution.  Spectra were then extracted by
summing rows at the location of the stellar signal, the number of which depends on the
seeing at the time of observation.  The two components of TWA 3, when observed 
simultaneously along the slit in 2002 December, 2003 February, 2009 December, and 2010 December
were well-separated.  Once extracted, the spectra were cleaned of bad pixels, flattened, normalized,
and corrected for barycentric motion. 
Figures~\ref{asa} and \ref{asb} show the final spectra from all our
observations of TWA 3A and 3B, respectively.  Effectively every feature in these
spectra is real; the signal-to-noise ratio was typically $>$200.
The double-lined nature of the TWA 3A spectra is prominent at most epochs. 
In the following we refer to the two components as TWA~3Aa and TWA~3Ab.
No line doubling is detectable in the spectra of TWA~3B.

\subsubsection{Imaging}

We observed TWA 3 in the $H$- and $K_S$-bands with the Clio imager at the Magellan 6.5\,m telescope using AO
on UT 2014 April 21.  Integration times were 0.1~s at $H$ and 0.3~s at $K_S$.
The images were flat fielded, and pairs of dithered images were
subtracted to remove the sky background and bad pixels.  As the TWA 3
A and B components were well-separated in the AO images, we used the
unresolved TWA 3A point spread function (PSF) to measure the relative
position and flux ratio of TWA 3B.  We used a plate scale of 15.846
$\pm$ 0.043 mas/pix and a rotation angle offset of -1.797$^\circ$
$\pm$ 0.159$^\circ$ (Morzinski et al. 2015).  Table 4
lists the separation, position angle, and flux ratios measured from these observations.
No evidence was seen indicating that the A and B components are themselves visual binaries.

\subsection{Optical}
\label{sec:optical_obs}

\subsubsection{Spectroscopy}

TWA 3 was placed on the observing program at the Harvard-Smithsonian
Center for Astrophysics (CfA) in 1998, along with other members of the
TW~Hya association. It was monitored between January of that year and
February of 2007 with two nearly identical echelle spectrographs
\citep[Digital Speedometers;][]{Latham:92} attached to the 1.5\,m
Tillinghast reflector at the F.\ L.\ Whipple Observatory (Mount
Hopkins, AZ) and to the 4.5\,m-equivalent Multiple Mirror Telescope
(also on Mount Hopkins), prior to its conversion to a 6.5\,m monolithic
mirror. A single echelle order 45\,\AA\ wide was recorded with
intensified Reticon photon-counting detectors, at a central wavelength
of about 5190\,\AA. This order contains the lines of the
\ion{Mg}{1}\,b triplet.  A total of 35 usable exposures of the star
were obtained at a resolving power of $R \approx
35,\!000$, with integration times ranging from 600 to 2400 seconds.
Given the northern location of these telescopes, all observations were
made at airmasses larger than about 2.8, and are rather weak: the
signal-to-noise ratios range from 7 to 15 per resolution element of
8.5~\kms. Nevertheless, as we show later this is sufficient to derive
meaningful RVs. The spectrograph slit was 1\arcsec, but
the seeing at low elevation was often very poor.  Consequently,
despite the angular separation of 1\farcs55 between TWA 3A and
TWA 3B, all of these spectra are likely to include light from both
visual components under typical seeing conditions.  Thorium-argon
spectra were obtained before and after each science exposure to set
the wavelength scale.  The zero point of the CfA velocity system was
monitored by taking exposures of the dusk and dawn sky, and small
run-to-run corrections for instrumental shifts were applied to the
velocities described later, following \cite{Latham:92}.

Additional optical spectra were collected with the fiber-fed FEROS
spectrograph mounted on the ESO 1.5\,m telescope (La Silla, Chile).
These observations cover the approximate wavelength range from 3600 to
9200\,\AA\ at a resolving power of 44,000. Integration times ranged
from 600 to 1800 seconds. We obtained a total of 8 usable exposures
between 1999 March and 2000 July, with signal-to-noise ratios of
25--35 per pixel around 5200\,\AA. Standard calibration frames
(Th-Ar-Ne lamp, flatfield, and bias exposures) were taken at the
beginning of each night, and the data reduction was performed under
MIDAS using the FEROS pipeline.  Telluric lines were used to
compensate for instrumental shifts, and additionally observations were
made each night of the standard star HR\,5777 \citep{Murdoch:93} to
check for RV shifts. These were found to be always smaller than about 30
m\,s$^{-1}$.  Because the spectrograph fiber has a 2\arcsec\ diameter,
all FEROS spectra also likely contain light of both visual
components of TWA 3.

Further observations of the star were obtained with the echelle
spectrograph on the 2.5\,m du Pont telescope at the Las Campanas
Observatory (Chile). Nine observations were gathered in 2006 from UT
January 30 to February 5. The wavelength coverage is 3500--10000\,\AA\
in 64 orders, and with a slit width of 1\arcsec\ the resolving power
is approximately 40,000. Exposure times were 900 seconds, and the
signal-to-noise ratios averaged 35--40 per pixel in order 32 centered
around 5170\,\AA.  Thorium-argon lamp exposures were taken before or
after each science exposure, and standard reductions were performed
with IRAF (bias subtraction, flatfielding). As with the CfA and FEROS spectra,
under typical seeing conditions most of the exposures probably include
light from both visual components of TWA 3, even though the slit used
was 1\arcsec\ wide.  On the first night, however, the seeing was good
enough to permit separate spectra of each star with little
contamination from the other.

\subsubsection{Imaging}

Optical imaging was carried out at Lowell Observatory's 4.3\,m
Discovery Channel Telescope (DCT) with the Large Monolithic Imager on UT 2014 April 8
using Johnson $B$ and $V$ and Cousins $R$ and $I$ filters; target exposure times were 30~s, 15~s, 5~s, and 1.25~s,
respectively.  Three 2$\times$2 binned images were taken of TWA 3 in each band.  Mean bias and flat
frames were created and applied to the individual target exposures.  We cross-registered images
in each individual band to align them and combined each set of three images, trimming the
final reduced output in each band to 3000$\times$3000 pixels for analysis.

Average seeing of $\sim0\farcs9$ facilitated
the extraction of photometry for the individual TWA 3A and B components.
Although without AO the individual PSFs were overlapping, we applied the
PSF of a nearby single star to model the relative positions and flux ratio of the pair using
techniques described in Schaefer et al. (2014).  The binned
plate scale\footnote{http://www2.lowell.edu/rsch/LMI/specs.html} was 0\farcs24 pix$^{-1}$.
Results of these observations are given in Table~\ref{tab.sepPA}.

\section{Radial Velocities}
\label{sec:rvs}

\subsection{Near-IR}
\label{sec:nir_rvs}

%In order to calculate a stellar RV, a standard star or synthetic spectrum is required as a template
%for cross-correlated against an observed stellar spectrum. In to measure the RVs of both stars in
%a spectroscopic binary, two templates
%are needed, one for the primary and one for the secondary, which are typically blended together in the
%observed spectrum.  A two-dimensional cross-correlation procedure is preferable
%(Zucker \& Mazeh 1994).  

%For our templates we used main sequence stars with spectral types similar to those

To measure RVs of TWA 3Aa and 3Ab we used a two-dimensional
cross-correlation code written at Lowell Observatory following the
TODCOR algorithm originally developed by Zucker \& Mazeh (1994). This
algorithm determines the individual RVs of both stars in the
spectroscopic binary simultaneously. It requires two templates, one
for each component, for which either synthetic or observed spectra may
be used. For this work we used observed spectra of main-sequence stars
of spectral types similar to those
of our T Tauri stars (Prato et al. 2002a; Bender et al. 2005). A nonlinear,
limb-darkened broadening kernel (e.g., Bender \& Simon 2008)
was applied to each template to mimic spectra with different rotational
velocities.  The spectral types, RVs, $v \sin i$, and $T_{\rm eff}$ values of the template stars we used to
optimize the cross-correlation with the TWA 3A IR spectra, GJ 402 and GJ 669B, appear in Table~\ref{spts}.
Templates broadened to $v \sin i$ values of 0$-$15~\kms\ were tested; we
found best fits using the M4 template GJ 402, broadened to a $v \sin i$ of 7~\kms\,, and the M4.5 template GJ 669B,
broadened to 5~\kms\,, for the spectra of TWA 3Aa and Ab, respectively.  
However, these values fall below our 2-pixel velocity resolution
element of $\sim$8~\kms, and thus we are limited to a general statement that the TWA 3Aa and Ab 
components have $v \sin i$ values less than 8~\kms~(but see \S 3.2).   For TWA 3B, we found a maximum
in the cross-correlation coefficient for a $v \sin i$ value of 15~\kms.
Given the high signal-to-noise spectra obtained in the IR and the low internal uncertainty in the
RV measurements of 0.2~\kms, we estimate that the largest uncertainty introduced into our
RV measurements of the TWA 3A components comes from the uncertainty in the template RVs
(Mazeh et al. 2003).

%We used a two-dimensional cross-correlation code written at
%Lowell Observatory following the TODCOR algorithm originally 
%developed by Zucker \& Mazeh (1994). This algorithm simultaneously determines the individual RVs
%of both stars in the spectroscopic binary and the correlation coefficient. 

% The component light
%intensity ratio may be left as a free parameter or may be set to a fixed value for all the
%input spectra for a particular binary. 

The TODCOR algorithm can also solve for the intensity ratio between the components.
The intensity ratio when the stars are at their largest RV separation gives the best results,
when available.  We first removed the two sets of points with RVs closest to the center of mass velocity for the system,
taken on UT 2005 February 22 (JD 2453423.9356) and UT 2010 December 12 (JD 2455543.1431), and left the
light ratio as a free parameter for the other 9 TWA 3A spectra.  We obtained an average light ratio of 0.61$\pm$0.11. 
This value was then fixed and the component RVs were then re-determined with TODCOR for all 11 spectra.
The resulting heliocentric RVs extracted from our IR spectra of TWA 3A using M4V and M4.5V template stars (Table ~\ref{spts})
are listed in Table 6.  The RVs for TWA 3B, determined from cross-correlation against GJ~15A (Table  ~\ref{spts})
also appear in Table 6 and show a RV range of 0.70~\kms.
Cross-correlating the strongest spectrum of TWA 3B from UT 2002 December 22 (Figure 2) against those from the three other epochs
indicated no RV shift greater than 0.3~\kms\ over a period of almost 8 years.  The standard deviation of the
TWA 3B $\Delta$(RV) measurements from these cross-correlations was 0.2~\kms, indicative
of our internal RV uncertainties for the IR spectroscopy of these bright target stars.

\subsection{Optical}
\label{sec:optical_rvs}

In view of the composite nature of our CfA spectra (TWA 3A $+$ TWA 3B),
and the knowledge from our near-IR observations that one of the
visual components is double-lined, we analyzed the CfA spectra using
TRICOR \citep{Zucker:95}, a three-dimensional extension of the
two-dimensional cross-correlation algorithm TODCOR \citep{Zucker:94}.
TRICOR uses three (possibly different) templates, one for each
star. We selected these templates from among a set of spectra
of M dwarfs taken with the same instrumentation as the target,
covering a wide range of spectral types (K7--M5.5).  Our choice was
guided by the spectral types of the best templates found from the
analysis of the near-IR spectra, and by extensive testing to identify the closest
match to the optical spectra as indicated by the highest
cross-correlation values averaged over all our CfA exposures. The best
compromise among the available templates yielded GJ~699
(Barnard's star) for the two
components of the double-lined binary TWA 3A and GJ~48
for TWA 3B, in good agreement with the spectral type standards used for
the IR analysis (Table~\ref{spts}).  Changing these templates by one subtype
resulted in relatively minor differences in the velocities, and did
not change our final solution significantly.  We added rotational
broadening to the templates with $v \sin i$ values of 7~\kms,
5~\kms, and 12~\kms\ for TWA 3Aa, Ab, and B, respectively, informed by
those used for the Keck spectra.  

Thirteen of our CfA observations were
contaminated by moonlight. In order to prevent biases in the
velocities, these spectra were analyzed with an extension of TRICOR to
four dimensions \citep[QUADCOR;][]{Torres:07}, selecting as the fourth
template a synthetic spectrum corresponding to the Sun, based on model
atmospheres by R.\ L.\ Kurucz.  We checked that in each case the
velocity for this fourth set of lines agreed with that expected from
the barycentric motion of the Earth. The measured CfA velocities for the
three components of TWA 3 transformed to the heliocentric frame are
reported in Table~\ref{tab:cfa_rvs}.  Typical uncertainties are about
2.7~\kms, 4.4~\kms, and 2.7~\kms\ for TWA 3\,Aa, Ab, and B,
respectively.  Because of significant differences in the line strengths in
each spectrum, the individual uncertainties take into account the
signal-to-noise ratio of each observation.  Following \cite{Torres:07}
we determined the light ratio between Ab and Aa to be $0.59 \pm 0.08$
at the mean wavelength of our observations (5190\,\AA). While in
principle the TRICOR/QUADCOR analysis can also provide the light ratio
between stars B and Aa, in practice this measurement is unreliable
because of slit losses.\footnote{The slit width is smaller than the
binary separation, resulting in varying amounts of light from stars A
and B entering the slit at each observation depending on guiding and
seeing.} The Ab/Aa light ratio is unaffected because the angular
separation between those stars is negligible compared to the slit
width.

Similar analysis techniques were applied to the FEROS and du Pont
spectra, using the same templates as above in view of the similar
resolving power of the instruments. For the du Pont spectrum in which
TWA 3A was observed separately from TWA 3B we applied TODCOR, as the
spectrum is only double-lined; for
TWA 3B alone, the RV was derived by one-dimensional
cross-correlation with the appropriate template.  The heliocentric
RVs from FEROS and du Pont are reported in
Table~\ref{tab:feros_rvs} and Table~\ref{tab:dupont_rvs},
respectively.  Typical uncertainties for FEROS are about 2.7~\kms\
and 3.7~\kms\ for the primary and secondary of TWA 3A, and 3.3~\kms\
for TWA 3B. For the du Pont spectra the errors are 1.5, 2.4, and 2.7~\kms,
respectively. The light ratios between stars Ab and Aa determined from
these spectra are $0.55 \pm 0.08$ for FEROS and $0.50 \pm 0.10$ for
du Pont at a mean wavelength of 5190\,\AA, consistent with the
measurement from the CfA spectra.  The result from the du Pont light ratios
is less reliable than the others because the du Pont observations were all made
at epochs when both components were close to the center-of-mass velocity.
The average of the CfA and FEROS light ratios, taken at favorable epochs,
is $0.57 \pm 0.06$.

\section{Orbital Solutions}
\label{sec:orbit}

\subsection{The TWA 3Aa--Ab Spectroscopic Binary}

Although the Ab/Aa mass ratio is relatively large ($q = 0.841$; see Table~\ref{tab:orbit}), early analysis of the optical data sets
did not yield a consistent solution, likely the result of the confusion introduced by the B component in the blended spectra (\S3.2).
Therefore, prior to combining all of the observations we carried out an
analysis of the 11 pairs of IR velocities alone (Table 6) using standard, nonlinear
Levenberg-Marquardt least-squares techniques (Press et al. 1992). This fit is shown in the second column of Table~\ref{tab:orbit}.
With this initial solution as a guide we were then able to obtain an independent solution
for the CfA optical data (Table 7), with weights for
the individual velocities inversely proportional to their uncertainties.
This CfA-only solution is shown in the third column of Table~\ref{tab:orbit},
%  We compared the
%independent orbital solutions to the extent possible as a check on
%possible systematic differences, particularly in the velocity amplitudes that determine the minimum masses.
%For both the Keck and CfA data sets the seven
%standard orbital elements were left free. There is good agreement
and shows good agreement with the previous fit, particularly for the velocity
semi-amplitudes that determine the minimum masses.
The only exception is the
center-of-mass velocity $\gamma$, for which the difference is likely the result of 
zero-point uncertainties discussed below.  Our du Pont spectra
(which were scheduled before we had determined the ephemeris for
TWA 3A) were obtained at very unfavorable orbital phases near
conjunction and do not allow for the calculation of an independent orbital solution.  The
phase coverage and number of the FEROS measurements are also
insufficient for a separate fit, but they do sample the velocity
extremes. A constrained solution using the FEROS data with the ephemeris and geometric
parameters ($e$ and $\omega$) held fixed from the Keck results yields
rough velocity amplitudes of $K_{\rm Aa} = 21.9 \pm 2.4~\kms$ and
$K_{\rm Ab} = 29.1 \pm 2.6~\kms$, consistent with those from the Keck
and CfA fits.

%For the final solution we combined the four data sets, and in order to
%allow for different velocity zero points, we solved for three velocity
%offsets between a reference data set and each of the other three
%groups.  CfA was chosen as the reference set, as it has the most
%measurements.  

\begin{figure}
\epsscale{0.95}
\plotone{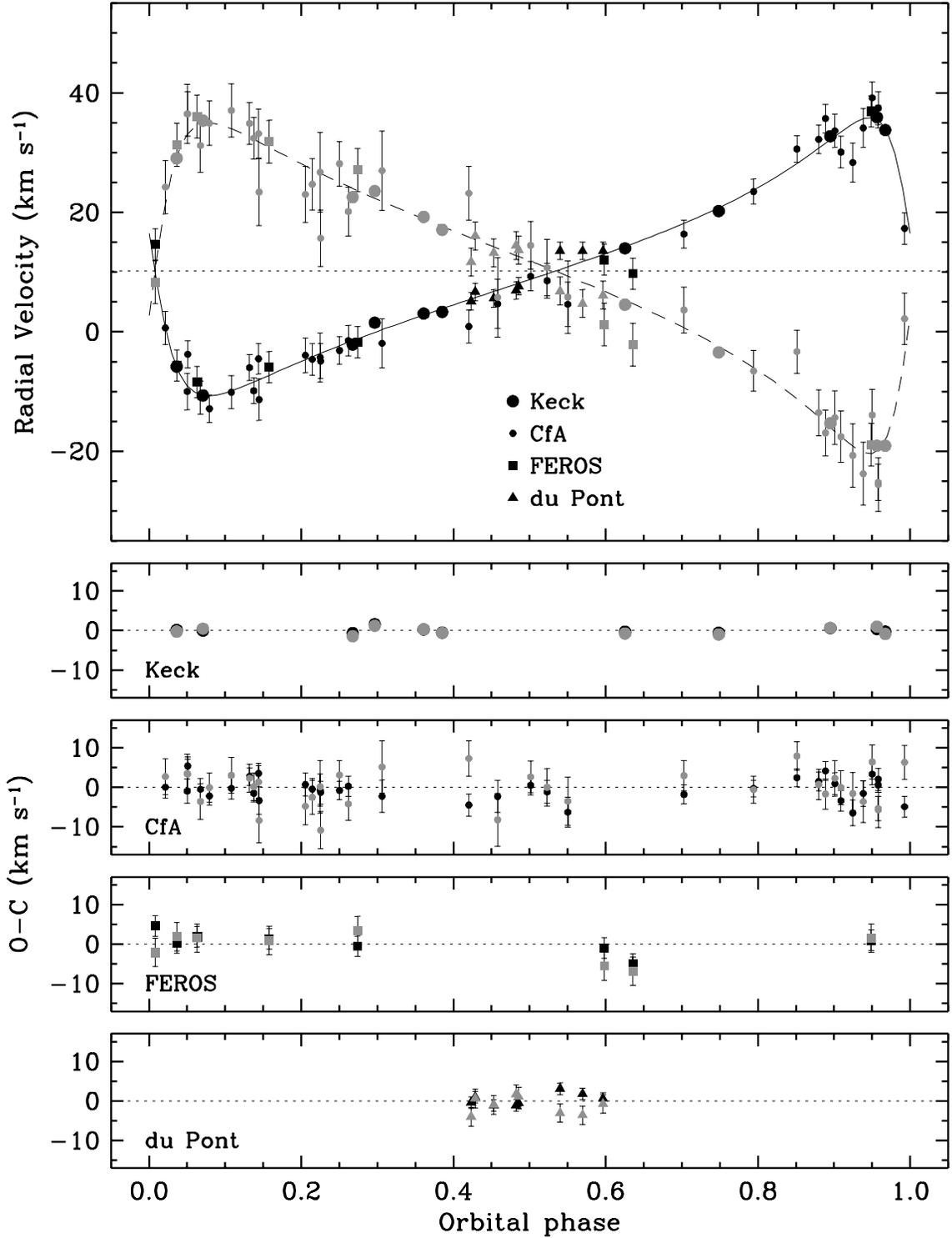}
%
%{\vskip -30pt} 
\caption{\emph{Top:} Radial velocities for TWA 3A
and our model from the best fit solution of Table~\ref{tab:orbit}
(solid line for the primary, dashed for the secondary). The dotted
line indicates the center-of-mass velocity of the system. Measurements
from different data sets are represented with different symbols, as
labeled. \emph{Bottom panels:} Velocity residuals ($O-C$) shown
separately for each data set with the same error bars as in the top panel.\label{fig:orbit}}
\end{figure}

For the final solution we combined the four data sets, and in order to
account for possible differences in velocity zero points we solved
for three offsets between each data set and the CfA set taken as the
reference, as it has the most measurements.
Additionally, because our analysis techniques do not
return internal errors for the velocities, we made initial estimates
of these for each data set based on the velocity scatter from
preliminary solutions, and then rescaled those errors by iterations in
our global fit so as to achieve reduced $\chi^2$ values of unity,
separately for each star and each data set. These final uncertainties
are the ones reported in
Tables~\ref{tab:keck_rvs}--\ref{tab:dupont_rvs}. The orbital elements
from our combined fit are listed in the last column of
Table~\ref{tab:orbit}, along with other quantities. This solution is
dominated by the near-IR velocities, which have the highest precision and
a weight that is more than 5 times greater than the next most precise
data set (du Pont). Nevertheless, the inclusion of the less precise
data has improved the formal uncertainties of all elements. A
graphical representation of our solution is shown in
Figure~\ref{fig:orbit}, along with residuals for each data set.

Of the three velocity offsets reported in Table~\ref{tab:orbit}, only
the one between CfA and Keck is statistically significant: $\Delta RV
({\rm CfA-Keck}) = +1.53 \pm 0.43~\kms$.  While this may be attributable in
part to instrumental effects, we note that the velocities derived from
the CfA, FEROS, and du Pont spectra (which seem to have consistent zero
points) all used the same templates, whereas those from Keck used
different ones. It is likely, therefore, that the zero-point
shift for Keck is the result of uncertainties in the RVs adopted for the templates.

We note that Malo et al. (2014) have recently published a few RV
measurements of all three visible components of TWA 3. Their
velocities for the B component agree well with ours, and although full
dates were not provided for their measurements of Aa and Ab, they too
seem consistent with our spectroscopic orbital solution.

\subsection{The Visual Orbit of the Close TWA 3Aa--Ab Binary}

\begin{figure}
\epsscale{0.8}
\plotone{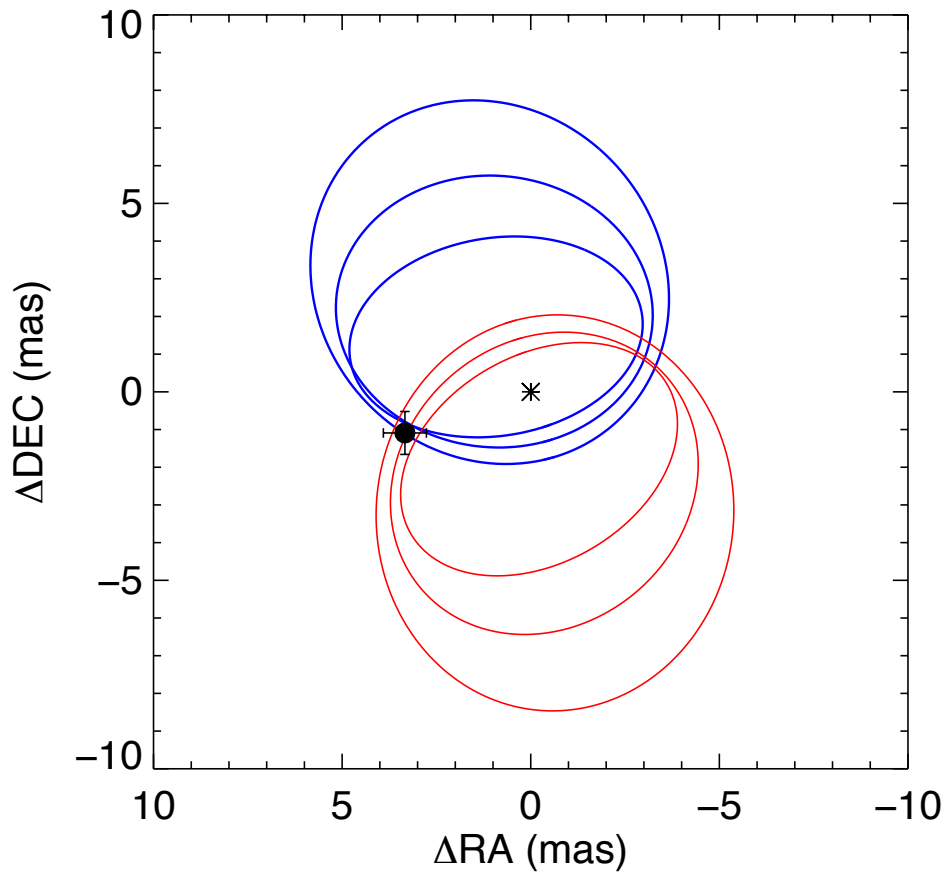}
\caption{Sample orbits consistent with the 1\,$\sigma$ uncertainties from the VLTI PIONIER separation measurement of TWA 3 Aa-Ab (filled black circle).  These orbits were found by fixing the spectroscopic orbital parameters and selecting values for $a$, $i$, and $\Omega$ at random.  If we add a constraint that the orbital parallax lies within $d = 34 \pm 4$~pc \citep{mamajek05}, then we find that the orbits cluster into two groups with $i < 90\arcdeg$ (red orbits) and $i > 90\arcdeg$ (blue orbits).}
\label{TWA3_AaAb}
\end{figure}

The close pair TWA 3 Aa--Ab was resolved spatially with the PIONIER combiner on the VLTI \citep{anthonioz15} at
a separation of $3.51 \pm 0.57$~mas and position angle of $108.1^\circ \pm 9.3^\circ$ on
HJD~2455601.870.  We explored the range of orbital solutions consistent with this
measurement by fixing the spectroscopic orbital parameters ($P, T, e, \omega$) and
selecting values for the angular semi-major axis, inclination, and longitude of the line
of nodes ($a, i, \Omega$) at random.  We found that the 1\,$\sigma$ confidence intervals
are well defined for $a$ (4.7--6.5 mas) and $\Omega$ ($93^\circ - 123^\circ$), but that the
confidence region for $i$ extends from $0^\circ - 180^\circ$.  If we add the constraint that the
orbital parallax (measured from the $K_1$, $K_2$, $P$, $a$, $e$, and $i$) lies within the uncertainties
of the parallax derived from the moving group cluster method \citep[$d = 34 \pm 4$~pc;][]{mamajek05},
then we find that the allowable ranges for $i$ fall into two families of solutions
between $32^\circ - 63^\circ$ and $118^\circ - 149^\circ$, as shown in Figure~\ref{TWA3_AaAb}.
These ranges yield absolute masses of $M_1$ from 0.17--0.86~$M_\odot$ and $M_2$ from 0.14--0.72~$M_\odot$.

Adopting a value for the angular semi-major axis, 5.6~mas, that falls in the average of the range of well-defined confidence intervals,
4.7--6.5 mas, and assuming a distance of 34 pc (Mamajek 2005), we estimate the physical size of the semi-major axis
to be $\sim$0.19~AU.  Given the eccentricity of the SB orbit, 0.628, we find an orbital
periastron distance of 0.07~AU and an apastron distance of 0.31~AU.

\subsection{The Visual Orbit of the Wide TWA 3A--B Binary}

We measured the relative RVs between the four epochs of observation of TWA 3B
shown in Figure 2 by cross-correlating the highest signal-to-noise spectrum, from UT 2002 Dec 22,
against all other epochs.  The RVs of TWA 3B appear constant, within the $\sim1~\kms$
uncertainties, thus there is no evidence to indicate that TWA 3B is also a spectroscopic binary.
The measured difference between the RV of TWA 3B and the center
of mass velocity of TWA 3A is $-1.21 \pm 0.25~\kms$ from the Keck observations
and $-0.73 \pm 0.60~\kms$ from the CfA velocities, consistent with a
physical association between the two visual components.

To investigate the properties of the TWA 3A--B orbit,
we downloaded measurements of the wide binary separation from the
Washington Double Star Catalog\footnote{http://ad.usno.navy.mil/wds} (Mason et al. 2001).
From this catalog, we used eight epochs from 1992 through 2015 to
investigate the relative motion between this pair of stars (Reipurth \&
Zinnecker 1993; Webb et al. 1999; Weintraub et al. 2000; Brandeker et
al. 2003; Correi et al. 2006; Mason et al. in prep; Janson et al.
2014; Tokovinin et al. 2015).  We supplemented these data with observations made at the DCT
(\S 2.2.2) and Magellan (\S 2.1.2) telescopes (Table 4).  For the
measurement reported by Reipurth \& Zinnecker (1993) we used an
observation date of 1992 Jan 8/9 (Besselian Year 1992.0216) supplied by B.
Reipurth (2015, priv. comm.).

The relative motion between TWA 3 A--B covers only a small arc (Figure 5).  We computed a linear least-squares fit to
measure the relative motion between the components in R.A. and
declination ($\Delta \mu_\alpha \cos{\delta} = 9.67 \pm 0.07$ mas~yr$^{-1}$, $\Delta \mu_\delta = -11.71 \pm 0.07$ mas~yr$^{-1}$).
We also computed binary orbit fits using the grid search procedure described in \citet{schaefer14},
but found that the $\chi^2$ from the orbit fit was indistinguishable from the linear fit.  
To determine realistic ranges for the orbital parameters, we randomly
searched the parameter space and added a constraint that the total
system mass, assuming a distance of 34 pc (Mamajek et al. 2005) must
be less than 2.0 $M_\odot$.  This is roughly double the expected total
mass of 0.8 $M_\odot$ based on the component spectral types and the 10
Myr evolutionary tracks computed from Baraffe et al. (2015).  Table 11
shows the ranges of orbital parameters for solutions obtained in our
random grid search procedure.  We show six example orbits in Figure 5
with masses in the range of 0.78--0.81 $M_\odot$ and inclinations
ranging from $120^\circ - 128^\circ$.  The parameters for the selected
orbits are given in Table 12; these orbits are not definitive, and
should only be used as a way to estimate the range of possible motion in the near future.

%Figure 6 shows that the relative linear motion between the TWA 3 A--B pair is smaller than the difference in
%proper motion between the unresolved TWA 3 system and other confirmed TW Hydra association members (Torres et al. 2003).  
%Given that the maximum $\Delta \mu$ described above is
%From the values for $\Delta \mu_\alpha$ and $\Delta \mu_\delta$ given above, we find a transverse motion of
%$\sim$15 mas~ yr$^{-1}$; we indicate the extent of this range in Figure 6 with a dashed circle of radius 15 mas~yr$^{-1}$ around TWA 3 at the
%origin of the plot.  For only one of the other cluster members is the $\Delta \mu$ as small as that between
%TWA 3A and 3B; for the majority of the members, the relative $\Delta \mu$ is much greater.
%We interpret this as suggestive that the TWA 3 A,B pair is likely bound.  Taken together with the stronger evidence from 
%the small difference in the TWA 3A--B RV quoted above, we believe that it is unlikely that TWA 3B is unbound.

The concordance of position, proper motion, and radial velocity all are consistent with TWA 3A and B being a bound pair.
To assess the likelihood that TWA 3A and 3B are gravitationally bound
to each other, we compared the total spatial velocity as determined
from radial velocity and proper motion measurements with the escape
velocity for TWA 3A. We estimated the combined mass of the TWA 3A
binary at 0.37~M$_{\odot}$, based on our spectral
classification. In the more conservative estimate, the difference in
RV between the 3A center of mass and 3B is $-1.21 \, {\rm km \,
s^{-1}}$. The total proper motion is $15 \, {\rm mas \, yr^{-1}}$
which equates to a space velocity of $2.4\, {\rm km \, s^{-1}}$, given
a total velocity difference of $2.7 \, {\rm km \, s^{-1}}$ between A
and B. The escape velocity from 3A is $\sqrt {2 G M_{\rm A} / r_{\rm
AB}}$, which equals $3.6 \, {\rm km \, s^{-1}}$ in the case where the
separation between the components is purely in the sky plane, $r_{\rm AB}
\sim 50 \, AU$. If we estimate an additional separation in the line of
sight direction equal to the sky plane separation, the escape velocity
drops to $3.0 \, {\rm km \, s^{-1}}$. In either case, 3A and 3B can be
plausibly argued to be bound to each other.

Finally, the HR Diagram displayed in Figure 6 shows that all three components are consistent with a 10 Myr age to within our uncertainties.  We used values for
$T_{\rm eff}$ determined for our best fit radial velocity template spectra (\S 3) by Mann et al. (2015).  The optical solutions used GJ 48 (M3) to fit the
B component and GJ 699 (M4) for both stars in the A component.  In the IR we found the best fits with GJ 15A (M3; Prato 2007) for the B component,
GJ 402 (M4) for the Aa component, and GJ 669B (M4.5) for the Ab component.  As both the IR and optical solutions used an M3 for
the B component, we determined the average $T_{\rm eff}$ for the two M3 stars with metallicity closest to solar given in Table 5 of Mann et al.
and propagated their uncertainties, yielding $T_{\rm eff}$(B)$=$3453 $\pm$86 K.  Although Mann et al. provide a $T_{\rm eff}$ estimate for
GJ 15A, they do not for GJ 48.  Prato (2007) indicates an M3 type for GJ 15A; however, Mann et al. found an earlier type of M1.4.
For the primary star in the spectroscopic binary, both the optical and IR solutions indicate an M4.  Mann et al. give $T_{\rm eff}=$3238 $\pm$60 K for
GJ 402, close to the 3228 $\pm$60 K value for GJ 699.  We used the former $T_{\rm eff}$.  For the spectroscopic binary secondary,
Mann et al. do not give an $T_{\rm eff}$ for GJ 669B so we again used the average of the two M4.5 stars in their Table 5 with metallicity
closest to solar and found $T_{\rm eff}=$3131 $\pm$85 K.  We calculated absolute H-band magnitudes using the A and B
component magnitudes from Webb et al. (1999) given in Table 13.  To determine the apparent H magnitudes of the Aa and Ab
components, we used the value for TWA3 A from Webb et al. in combination with the average spectroscopic binary
H-band flux ratio, 0.61 $\pm$0.11, found from cross-correlation
of 9 epochs of IR spectroscopy (\S 3.1).  We measured the absolute $H$-band magnitudes using $d = 34 \pm4$~pc (Mamajek 2005).  The coevality
of the three components does not necessarily imply that they formed together but the lack of a discrepancy
in age is in any case a requirement for a common origin.

\section{The Distribution of Stars, Gas, and Dust in the TWA 3 System}

TWA 3 has been historically characterized as a classical T Tauri, i.e., a disk-bearing system.  De la Reza et al. (1989)
identified the wide A--B binary for the first time ``at the Coude focus" when they first observed TWA 3.
They determined that at least one of the TWA 3 components was associated with
the coincident IRAS source, implying the presence of warm dust, and found H$\alpha$ equivalent widths of 
20\,\AA\ and 8\,\AA\ for the A and B components, respectively, which they interpreted to imply that both A and B
were actively accreting circumstellar gas; this interpretation has not been born out for the B component in subsequent observations
and it appears to be diskless.
Webb et al. (1999) comment in their Table 1 that TWA 3A is a candidate spectroscopic binary; Muzerolle et al. (2000) and
Torres et al. (2003) provided additional evidence for this short-period pair.  With our subsequent
high-resolution IR spectroscopy campaign we demonstrate
that TWA 3A, the ``primary" star, is actually two later-type stars in a 35 day period orbit (\S4.1).
Jayawardhana et al. (1999) showed that the mid-IR excess, implying the presence of warm dust, is attributable to only the A component.
Andrews et al. (2010) resolved the TWA 3A circumbinary disk at submillimeter wavelengths.
We present evidence for an accretion disk in the TWA 3 system in the form of updated SEDs
and hydrogen emission line equivalent widths, demonstrate that the circumbinary disk around TWA 3A
is not aligned with either stellar binary orbital planes, and examine the possible capture of TWA 3B as a potential formation mechanism for this system.

\subsection{SEDs and Accretion: Dust and Gas in TWA 3}

Figure 7 shows the individual SEDs for TWA 3A and TWA 3B.  Although SEDs for this system have appeared elsewhere
(e.g., Andrews et al. 2010) we provide specific literature sources and fluxes (Table 13) for future reference.
Zuckerman (2001) suggested that the entire TWA 3 system was surrounded by cool, large grains; however, Andrews et al. 
found no evidence for this material.  All excess IR and submillimeter flux is localized in a dusty circumbinary gas disk around
the TWA 3A component only, indicated by the mid-IR and longer wavelength points in Figure 7.  TWA 3B was either not
detected at longer wavelengths or the two components were too blended for individual flux densities to be determined.
However, given that the one upper limit given for the B component at 10 $\mu$m is 18$\times$ fainter than the flux density of
TWA 3A at that wavelength, it is safe to assume that the vast bulk of the longer wavelength emission originates in the TWA 3A circumbinary disk.

Andrews et al. (2010) modeled the resolved
submillimeter flux and provided best fit parameters for the disk, including an outer disk radius of 15--25 AU
and an inner radius for the dense disk of 1 AU.  In their model, material 500$\times$ more tenuous extends down to 0.4 AU.
The TWA 3Aa--Ab probable family of orbits described in \S4.2 and shown in Figure 4 has a range of semi-major axes of
4.7--6.5 milliarcseconds.  For an average value of 5.6 mas, and the distance to TWA 3 of 34~pc, the corresponding physical
separation of 0.19 AU, is consistent with the clearance of a small central hole in the disk model described by Andrews et al. (2010).
Muzerolle et al. (2000) describe their unambiguous detection of active gas accretion onto at least one star in the TWA 3Aa--Ab
pair, although they calculate a relatively low mass accretion rate of 5$\times$10$^{-11}$ M$_{\odot}$~yr$^{-1}$.

We fit our SEDs with the solar metallicity (Fe/H within $\pm$ 0.05)
spectral templates from Mann et al. (2015) using both our broad and
narrow-band photometry (Table 13) taken at $\lambda < 3.0 \mu$m. For
the $I_C$ and $R_C$ filters, we averaged the values for the angularly
unresolved TWA 3 system given in Table 13 and used the A/B flux ratios
in Table 4 to calculate the individual A and B component magnitudes
and corresponding fluxes. Fixing $A_V$ to a value of 0.01 mag
(McJunkin et al. 2014) and scaling the flux of the spectral templates
to match the observed fluxes resulted in a best fit for TWA 3A to the
M4.1 template (PM I19321-1119) and for TWA 3B to the M3.5 template (PM
I09553-2715), consistent with our optimal cross-correlation results.
At a distance of 34 pc, summing the flux under the model SED
(using the Rayleigh Jeans tail of a blackbody curve to approximate the
flux at wavelengths longer than 3 $\mu$m) yields luminosities of
$0.081~\pm0.003~\pm0.022~L_\odot$ and $0.055~\pm0.002~\pm0.015~L_\odot$ for the A and B components, respectively.   The first
uncertainty is computed from the SED fit while the second is from
propagating the $\pm 4$ pc uncertainty in the distance.

\subsection{Disk-Orbit Alignment}

According to the model of Andrews et al. (2010), the TWA 3A circmbinary disk inclination is 36\arcdeg~and the position
angle is 169\arcdeg\ east of north.  In Figure 8 we show a schematic of this model disk with the family of
probable orbital fits to the arc of observed positions (Figure 5).  In order to calculate the relative inclination between
the disk and the outer orbit of TWA 3B, we followed the approach of Fekel (1981) for calculating the relative orbital
inclination $\phi$ in a hierarchical triple by substituting the disk's position angle and inclination for the inner orbit and a typical orbit from those
shown in Figure 8 for the outer orbit ($P=671.73$ years, $T=2451805.0$, $e=0.3976$, $a=2.100''$ or 71.40~AU at $d=34$~pc, $i=124.91\arcdeg$,
$\Omega=133.24\arcdeg$, $\omega=133.43\arcdeg$, $M_{total}=0.807~M_{\sun}$).  The disk's position angle is that of the line of
nodes and is not necessarily the position angle of the {\it ascending} node $\Omega_{disk}$, required by the Fekel formula, because it is
not possible to distinguish the ascending node from the descending node without RVs for the disk, which have not been measured.
Because of this 180\arcdeg\ ambiguity in $\Omega_{disk}$, the calculation of $\phi$
yields two results; for the $+$ sign in Fekel's equation we obtain $\phi=$86\arcdeg\ and for the $-$ sign we find $\phi=$32\arcdeg.  This analysis
suggests a significant discrepancy between the disk inclination and wide orbit inclination; it is even possible that these are close to perpendicular.

We applied the same experiment to the disk with respect to the inner binary.  Restricting the orbital parallax to be within 34$\pm$4~pc, we determined
the median inclinations and values for $\Omega$ along with their associated 1$\sigma$ uncertainties for the two families of orbits shown in Figure 4:
$i=136\arcdeg ^{+13\arcdeg}_{-18\arcdeg}$, $\Omega=103.4\arcdeg \pm6.4$ and $i=43\arcdeg ^{+20\arcdeg}_{-12\arcdeg}$, $\Omega=112.3\arcdeg \pm6.6$.
For the former inner orbit family, the $+$ sign in Fekel's equation yielded $\phi=$139\arcdeg\ and the $-$ sign $\phi=$114\arcdeg.
Even with uncertainties on the order of $\pm$20\arcdeg\, this implies a highly misaligned and retrograde orbit.
For the latter inner orbit family, the $+$ sign in Fekel's equation yielded $\phi=$36\arcdeg\ and the $-$ sign $\phi=$68\arcdeg\,
implying that a significant misalignment is also possible for these values.
We can also use the range of values for the semi-major axes in the two families of
orbits consistent with the VLBI measurement shown in Figure 4, $a=6.5$--4.7~mas, together with the value for  $a \sin i$ given in Table~\ref{tab:orbit} for the
combined IR and visible light spectroscopic binary solution, 27.34~$R_{\odot}$, assuming a distance of 34~pc.  For $a=6.5$~mas we obtain an inclination
of 35\arcdeg\ and for $a=4.7$~mas an inclination of 53\arcdeg, a similar range as described in the latter calculation above.

In Figure 8 we show the typical inner orbits as in Figure 4 with a blow up
of the inner circumbinary disk.  It appears likely that neither the inner spectroscopic binary orbit  nor the outer visual binary orbit is
co-planar with the disk.  In contrast, Czekala et al. (2016) found that the $\sim$16 day period DQ Tau spectroscopic binary is aligned with its
circumbinary disk to within 3 degrees at the 3$\sigma$ level.  This discrepancy in the TWA 3 system may be related to the presence of a torque on the disk
originating in the action of the tertiary.

\subsection{Origin of the TWA 3 System}

The TWA 3 system presents a counterintuitive example of a young hierarchical triple with no evidence for accretion disk material around
the single star in the wide, several hundred year orbit, but an actively accreting circumbinary disk surrounding the two stars in
a $\sim$35~day spectroscopic binary orbit.  Given its disk characteristics, TWA 3A falls into the rare class of double-lined, classical T Tauri 
spectroscopic binaries along with other examples such as DQ Tau (Czekala et al. 2016), V4046 Sgr (Rosenfeld et al. 2012), and UZ Tau E (Prato et al. 2004;
Mart\'{i}n et al. 2005).  Uniquely, however, TWA 3 is the lowest mass system in this group, and one of the oldest.
Furthermore, not only is the TWA 3A spectroscopic binary surrounded by an actively accreting
disk, it is also accompanied by a tertiary low-mass stellar companion which lacks any evidence of disk material.
If the three stellar components in this system formed concurrently, we are left with the question of why only
the close binary hosts a disk.  One obvious advantage is the higher mass of the double star in terms of gravitational potential, but
this is offset by the disadvantage of the dynamical impact on disk material by the orbital action of the close binary.

Considering the available evidence to the effect that the A--B pair appears bound (\S 4.3), it is attractive to ascribe the mismatched disk
properties of these wide components to the result of a dynamical capture event.  Clarke \& Pringle (1991a), however, show that
capture rates in the known star forming regions, even of a star by a massive disk, are too low to provide an important binary formation
mechanism.  The disk surrounding TWA 3A has below average dust mass at 7$\times$10$^{-6}$ M$_{\odot}$ (Andrews et al. 2010), although it was
likely more massive when the system was newly formed.  Very wide pairs (separation $>$1000 AU) can also form
though capture in the late stages of cluster dispersal (e.g., Kouwenhoven et al. 2010); the probable family of orbits shown in Figure 5,
however, possesses semi-major axes between 70 and 80 AU, too tight for the soft binaries formed as per Kouwenhoven et al. during
cluster dissolution.  Alternatively, the TWA 3 system could have been a loose aggregate of stars which was subsequently reconfigured
as a tight binary and a relatively wide tertiary (e.g., Reipurth \& Mikkola 2012).  In this case, the closer binary's interaction with the
disk material would tighten the orbit and the TWA 3B tertiary would be isolated outside of the binary$+$disk system.  However,
the N-body simulations of Reipurth \& Mikkola treat the evolution of compact triple systems which evolve to a configuration with
a tight pair and an extremely distant, i.e. hundreds of AU, tertiary, a scale far greater than that we observe in the TWA 3 triple.

Kaplan et al. (2012) examined the probability of a purely gravitational interaction between a close brown dwarf
binary and a solar type star.  Although TWA 3 is composed simply of 3 low-mass stars, we explored the results of their work in
search of clues to possible formation mechanisms. The likelihood of capture in the brown dwarf pair -- solar type star
model is $\sim$0.1\% and thus unlikely.  However, given a low-mass tertiary, the possibility of capture could be higher, although
it seems unlikely.

\section{Discussion}

In spite of the nominal age of 10 Myr for the TW Hya association (e.g., Webb et al. 1999), consistent with our
results (Figure~6), on-going gas accretion
surprisingly points to an actively accreting circumbinary disk around TWA 3A.  The eccentric binary is expected to perturb
such a disk, clearing out a hole in the center; Aguilar et al. (2008) find that the
higher the eccentricity of a system, the larger the inner radius of the circumbinary disk.  Figure 8 shows possible
visual orbits of the spectroscopic binary compared to the size of the tenuous inner disk, 0.4--1.0~AU, as modeled
by Andrews et al. (2010).  Given the large eccentricity of the spectroscopic binary of $\sim$0.63, this tenuous gas disk
cannot be stable under the criteria of Artymowicz \& Lubow (1994) and likely represents the streaming gas from the
disk outside of 1 AU to the accretion point with the stars, just inside of 0.4 AU.
Yang et al. (2012) measured the FUV emission in TWA 3A and B and detected H$_2$ emission lines
and a continuum excess in the A component but not in B.  Although they point out that the low Ca II/C IV luminosity ratio of
TWA 3A could result from purely chromospheric activity, the strong H$\alpha$ emission, H$_2$ emission, and
Balmer continuum emission all point to active accretion from the TWA 3A circumbinary disk onto one or both of the central stars.  
Herczeg et al. (2009) report the presence of [OI] emission around TWA 3A as well, indicative of outflow.

The lack of near-IR excess emission
reflects the destruction of the inner dust disk by the action of the spectroscopic binary orbit, leaving only
the tenuous gas (Andrews et al. 2010) detected by Yang et al. (2012) as hot H$_2$ and by Webb et al. (1999),
Muzerolle et al. (2000), and Herczeg et al. (2009) in the H$\alpha$ emission line.  Interestingly, the
crystalline silicate emission detected by Uchida et al. (2004) points to significant thermal processing in 
the dust outside the central cavity, delineated by the modeling of Andrews et al. (2010) at $\sim$1 AU.
Herczeg \& Hillenbrand (2014) determine a relatively young age of 3--4 Myr for the TWA 3 system, in
contrast to the canonical $\sim$10 Myr for TW Hya association stars and consistent with our results.

Although a number of very young systems such as UZ Tau E (Prato et al. 2002b), DQ Tau
(Basri et al. 1997), and AS 205B (Eisner et al. 2005) possess actively accreting, circumbinary 
disks, these systems have ages of just 1 to a few Myr.  Other components in these systems,
such as UZ Tau W and AS 205A, also boast active, optically thick disks.  What challenges our 
understanding of disk evolution in the much older TWA 3 system is the unexpected presence of an active disk around 
the close binary rather than around the single visual companion. It is possible, given
the relatively short period and high eccentricity, and the presence of TWA 3B at only
$\sim$70 AU, that the eccentric spectroscopic orbit of TWA 3A may have been impacted by the Kozai pumping
induced by TWA 3B.  Another 10 Myr old system, TWA 4 (HD 98800), is similarly puzzling:
a quadruple system in which the visual companions
are themselves spectroscopic binaries (Torres et al. 1995), with evidence for a
circumbinary disk only around the visual secondary component, TWA 4B (e.g., Prato et al. 2001). As Torres et al. showed, the
TWA 4B pair has a high eccentricity, $\sim$0.8, but the diskless single-lined binary, TWA 4A also
has a relatively high eccentricity of $\sim$0.5.

It has been speculated that the existence of a companion will prolong the depletion of the
circumstellar disk and slow the rotation of the stars (Armitage \& Clarke 1996).
If this is correct, then it might explain why there is still a disk present in the system but
not why it is around the spectroscopic binary component of the system rather than the single
star component. Bouvier et al. (1997) and White \& Hillenbrand (2005) suggested that disk truncation 
and tidal interactions inhibit accretion flow which would thus increase the lifetime of this disk as well. 
This theory is contrary to evidence suggesting that the circumbinary disk around TWA 3A is still accreting, however.
Lubow \& Artymowicz (1997) showed that if the internal gas pressure in the circumbinary disk is large enough to
overcome the energy barrier of the resonance, inward gas streams could flow across the gap onto the central
stars. This could then explain why the system is still accreting but it can not explain the lifetime of the disk.

If indeed TWA 3A and 3B are bound and both formed with disks, we speculate that the same sequence of tidal interactions 
invoked by Prato et al. (2001) may have disrupted the short-timescale dispersal of the TWA 3B disk.  Prato et al. studied the properties
of the quadruple TWA 4 and hypothesized that both of the tight, spectroscopic pairs in that system may have originally
hosted circumbinary disks, but that the difference in relative disk inclinations may gave ripped the circumbinary disk from TWA 4A early
in the formation history of the system.  Given the apparent lack of alignment, however, between either of the stellar binary orbits and the
disk plane, justification for the survival of the lone circumbinary disk is not obvious.  At best we might evoke a particular
set of parameters which ensure the stability and contribute to the longevity even of the remaining disk.

Future observations which could ultimately help define the system are straightforward for the inner orbit and circumbinary disk:
even a small number of additional VLBI visibilities will facilitate measurement of the visual orbit and hence individual
masses for the 35-day period binary.  ALMA observations of the disk will refine its inclination and $\Omega$ values, and
at high spatial resolution may reveal structure and detail both within the disk and of the outer radius.
Although numerous historical plates of the TWA 3 system have been digitized, the extremely poor image resolution makes analysis
of the relative positions of at 1\farcs.55 binary impossible.  To realize a more definitive TWA 3A--B orbit it will likely be
necessary to wait another 20 or 40 years.

\section{Summary}

We report a $\sim$35 day period spectroscopic companion to the visual primary star
in the TWA 3 system.  We find that the primary component (Aa) of the spectroscopic binary
has an M4 spectral type with a $v \sin i$ of $\sim$7~\kms\ and the secondary
component (Ab) has an M4.5 spectral type with a $v \sin i$ of $\sim$5~\kms. 
As expected for these spectral types, the two stars in the spectroscopic binary have a relatively high mass
ratio, 0.841 $\pm$ 0.014. TWA 3B, the visual companion at 1\farcs55, has an M3.5 spectral type with
a $v \sin i$ of $\sim$12~\kms and is apparently single.  Values of $v \sin i$ given here are taken from the optical data sets (\S 3.2).
Based on all available data, from our own spectroscopic solution for TWA 3A as well as multiple sources from the literature,
we have determined that the one disk present in the TWA 3 system orbits only the spectroscopic pair and is not aligned
with either the short-period inner or long period outer orbit.  We speculate that the longevity and apparent stability
of this disk is related to the unusual dynamics in this complex system but we are at a loss to explain the
lack of any circumstellar material orbiting TWA 3B, if indeed this component formed with a disk.
All indications, relative motion, age, and formation hypotheses, are consistent with the TWA 3A and 3B components
being stably bound.  TWA 3 shares the multiple star complexity and disk longevity with several other prominent
systems in the TW Hya region such as TWA 4, TWA 1, and TWA 5.  We conjecture that relatively unusual initial
conditions in this small association contributed not only to the high multiplicity fraction of the members but also
possibly to the long-lived and unusual disk configurations.  These properties may also be related at least in part
to an observational bias in that the TW Hya association is located at $<$50 pc, allowing detailed study not only of
the disk properties but also of both visual as well as spectroscopic binaries.

%We find that the system TWA 3 is a triple system with a visual primary spectroscopic
%binary and a visual secondary single star. Our analysis reveals that the spectroscopic
%binary has a period of 34.68 days, a center-of-mass velocity of 8.38 km s$^{-1}$, a mass
%ratio of 0.786, and an eccentricity of 0.613. 
%Based on the semi-major axis, we calculated the inner radius of the circumbinary disk to
%be $\sim$0.627 AU based on the $\sim$3$a$ value given by Artymowicz \& Lubow (1994). This value is a bit smaller than the previous calculations by Andrews et al. %(2010) and Uchida et al. (2004) but we have the advantage of knowing the orbital period. The projected separation of $\sim$70 AU of the visual binary components
%gives the outer edge of the circumbinary disk a radius of approximately 23 AU (1/3 projected separation; Artymowicz \& Lubow 1994).

%Given the multiplicity and age of the system, we would not expect to find any type of
%circumstellar material but if there was material, we would expect to find it around the
%star without a spectroscopic companion. However, this is not the case. The circumstellar
%material is actually found around the spectroscopic binary star making it circumbinary
%material. Not only is this disk present, it is still accreting. It is difficult to understand
%why this is the case.  To resolve the mystery we likely require a much larger sample
%of similar systems in addition to developing more nuanced and robust models and
%simulations of multiple star $+$ disk formation and evolution.

\acknowledgments

We thank numerous observatory TOs, OAs, and SAs for their exceptional observing support over
the many years of data collection for this project.  LP thanks Otto Franz, Cathie Clarke, and Steve Lubow,
among others, for helpful discussions which improved this manuscript.
We are grateful to P.\ Berlind, M.\ Calkins, G.\ Esquerdo, D.\ Latham,
and R.\ Stefanik for help in obtaining the CfA observations of TWA 3,
and to R.\ J.\ Davis for maintaining the CfA echelle data base over
the years.  LP thanks AB for inspiring discussions during the
completion of this work.  We are extremely grateful to the rapid response of the referee and to
the referee's comments and suggestions which have improved this manuscript.
LP, KK, IA, LHW, and DRR were supported in part by NSF grants AST-1009136 and AST-1313399 (to LP).
Initial work on this project by KK and IA was also supported by NASA Space Grant, through Northern Arizona University.
GT acknowledges partial support for this work from NSF grant AST-1509375.  GHS acknowledges support
from NSF grant AST-1411654.  
KMM's and LMC's work was supported by the NASA Exoplanets Research Program (XRP) by cooperative agreement NNX16AD44G.
These results made use of the Discovery Channel Telescope at Lowell
Observatory. Lowell is a private, non-profit institution dedicated
to astrophysical research and public appreciation of astronomy and
operates the DCT in partnership with Boston University, the
University of Maryland, the University of Toledo, Northern Arizona
University, and Yale University. The Large Monolithic Imager was
built by Lowell Observatory using funds provided by the National Science Foundation (AST-1005313).
We thank Sean Andrews for providing us with the 880 $\mu$m dust continuum contours from Andrews et al. (2010)
and Bo Reipurth for locating the exact date of his first measurement of the TWA 3A--B pair shown in Reipurth \& Zinnecker (1993).
We are grateful to T. Boyajian for a useful discussion on the best
templates to use in the SED fits and to A. Mann for providing the spectral templates.
This research has made use of the Washington Double Star Catalog maintained at the U.S. Naval Observatory and the SIMBAD reference database, the NASA Astrophysics Data System, and the data products from the Two Micron All Sky Survey, which is a joint project of the University of Massachusetts and the Infrared Processing and Analysis Center/California Institute of Technology, funded by the National Aeronautics and Space Administration and the National Science Foundation. Data presented herein were obtained at the W. M. Keck Observatory from telescope time allocated to the National Aeronautics and Space Administration through the agencyÕs scientific partnership with the California Institute of Technology and the University of California. The Observatory was made possible by the generous financial support of the W. M. Keck Foundation. We recognize and acknowledge the significant cultural role that the summit of Mauna Kea plays within the indigenous Hawaiian community and are grateful for the opportunity to conduct observations from this special mountain.

\clearpage

\clearpage

\begin{figure}
\centering
\includegraphics[width=6.0in]{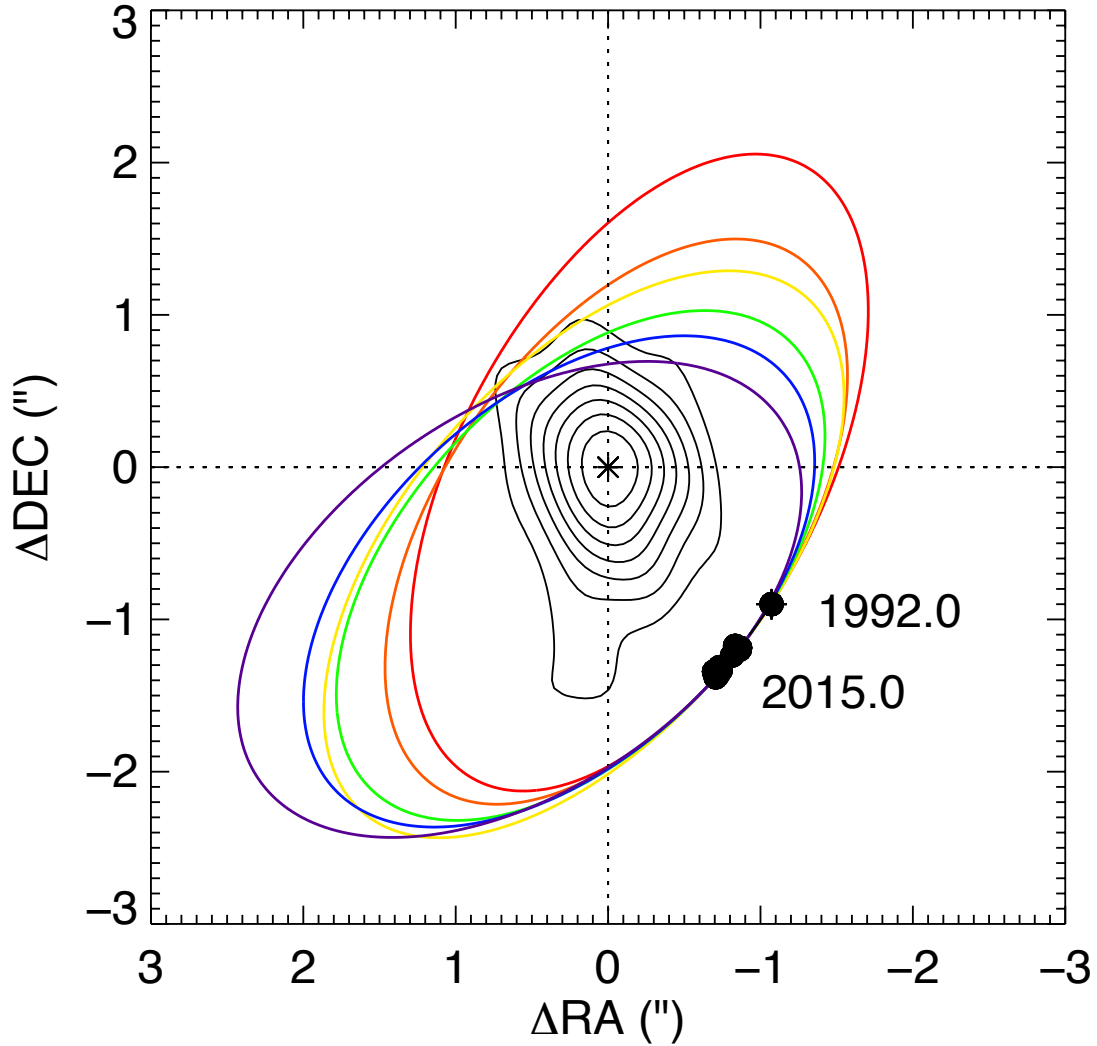}
\caption{Motion of TWA 3B relative
to TWA 3A from observations listed in the Washington Double Star
Catalog and the Magellan and DCT observations.  The rainbow lines show
examples of orbits that fit the data and produce a total mass in the
range of $0.79 - 0.81 M_\odot$, assuming a distance of 34 pc.  The
parameters for the plotted orbits are listed in Table 12.  The contours show the 880 $\mu$m
dust continuum emission from the disk around TWA 3A, convolved with the beam size.  These are plotted in 2$\sigma$ intervals,
beginning at 3$\sigma$ (10 mJy beam$^{-1}$; from Andrews et al. 2010).}
\label{orbs+disk}
\end{figure}

\clearpage

%\begin{figure}
%\centering
%\includegraphics[width=5.5in]{fig6.pdf}
%\caption{Relative proper motions of TWA 3 and other association members confirmed by Torres et al. (2003).  The dashed circle delineates a
%15 mas yr$^{-1}$ radius around the red square at [0,0], consistent with the motion of TWA 3B relative to that of TWA 3A.  Black triangles show the
%difference in proper motions between TW Hya members as given by Torres et al. (2003) and TWA 3, demonstrating that only one other star in
%the association has a proper motion difference on the order of that between TWA 3A and B, suggesting that TWA 3A and B are a bound pair.}
%\label{pms}
%\end{figure}

%\clearpage

%\begin{figure}
%\centering
%\includegraphics[width=5.0in]{dartmouth2.pdf}

\begin{figure}
\epsscale{0.8}
\plotone{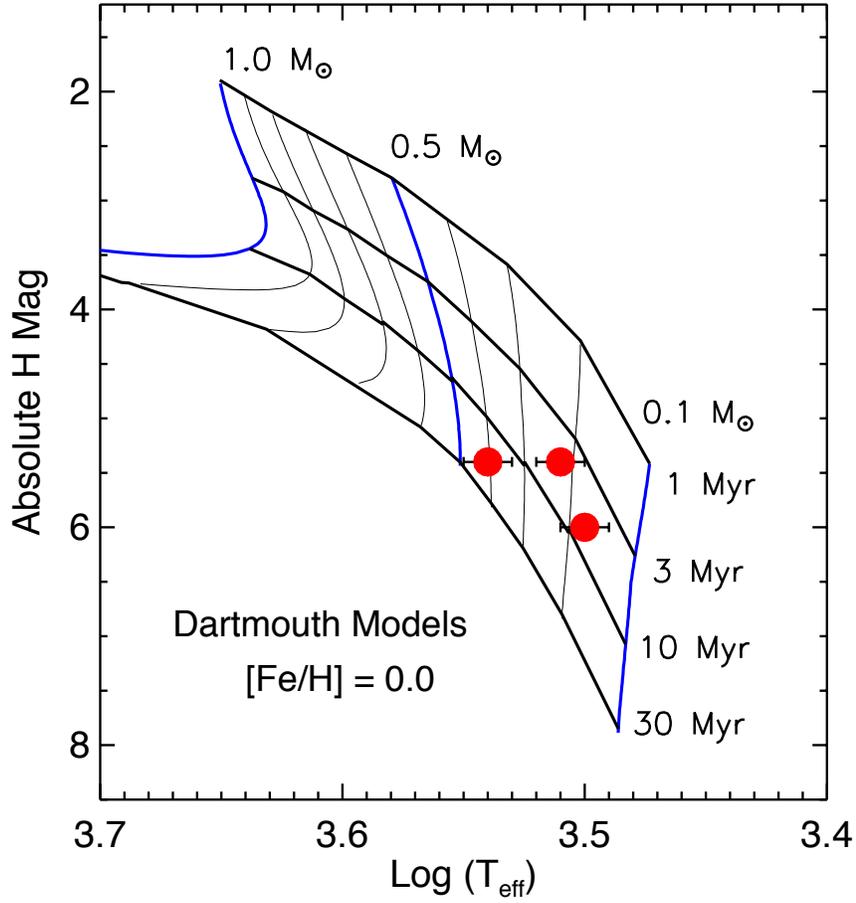}
\caption{H-R diagram for all three TWA 3 components.  Mass tracks and isochrones of the solar metallicity
Dartmouth evolutionary tracks (Dotter et al. 2008) are labeled.  See text for further details.  
The uncertainties in absolute H magnitude are all 0.1 mag, smaller than the plotted points.
All three components' positions are consistent with an age of 10 Myr.  \label{fig:hrd}}
\label{pms}
\end{figure}

\clearpage

\begin{figure}
\epsscale{2.0}
\plottwo{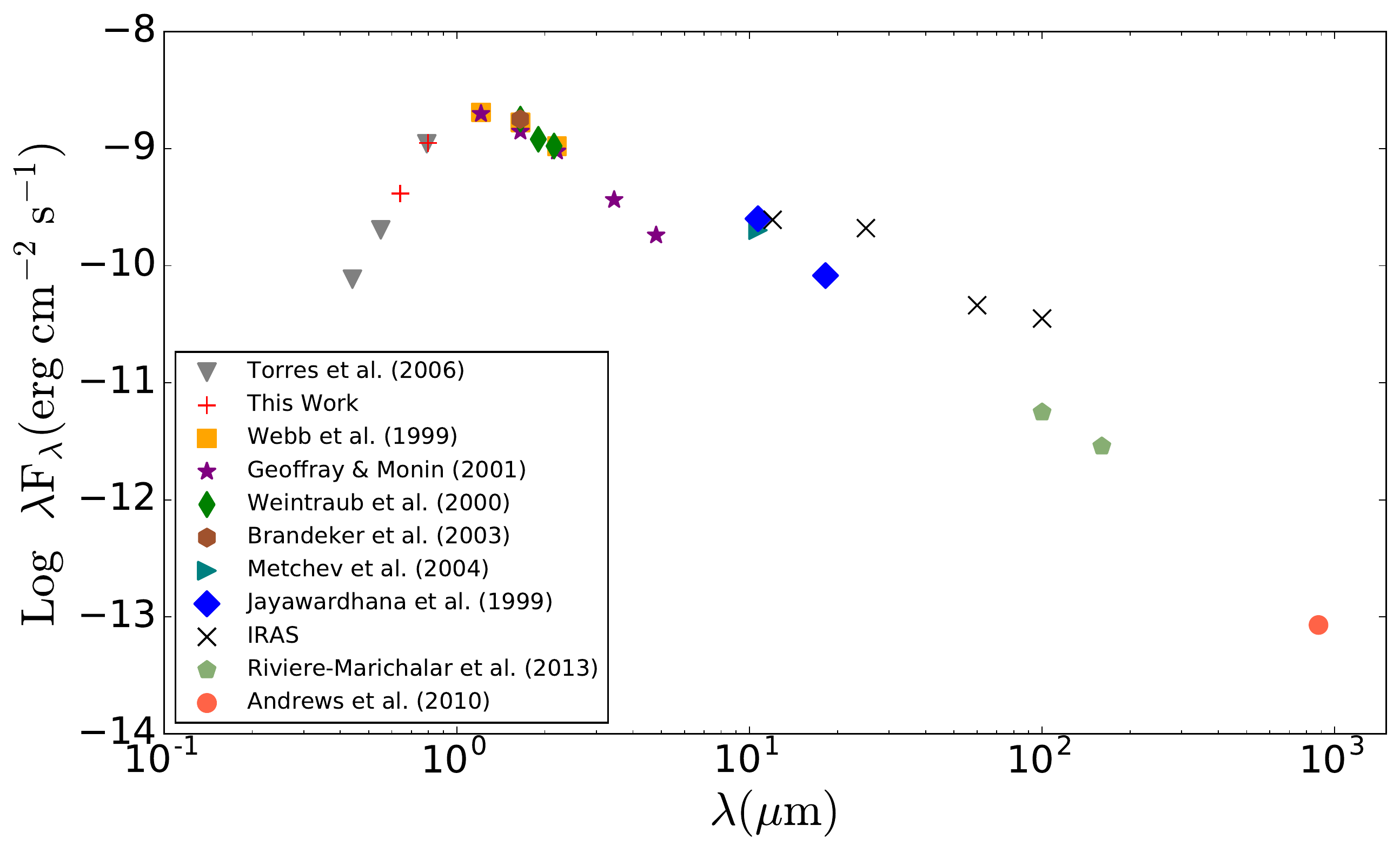}{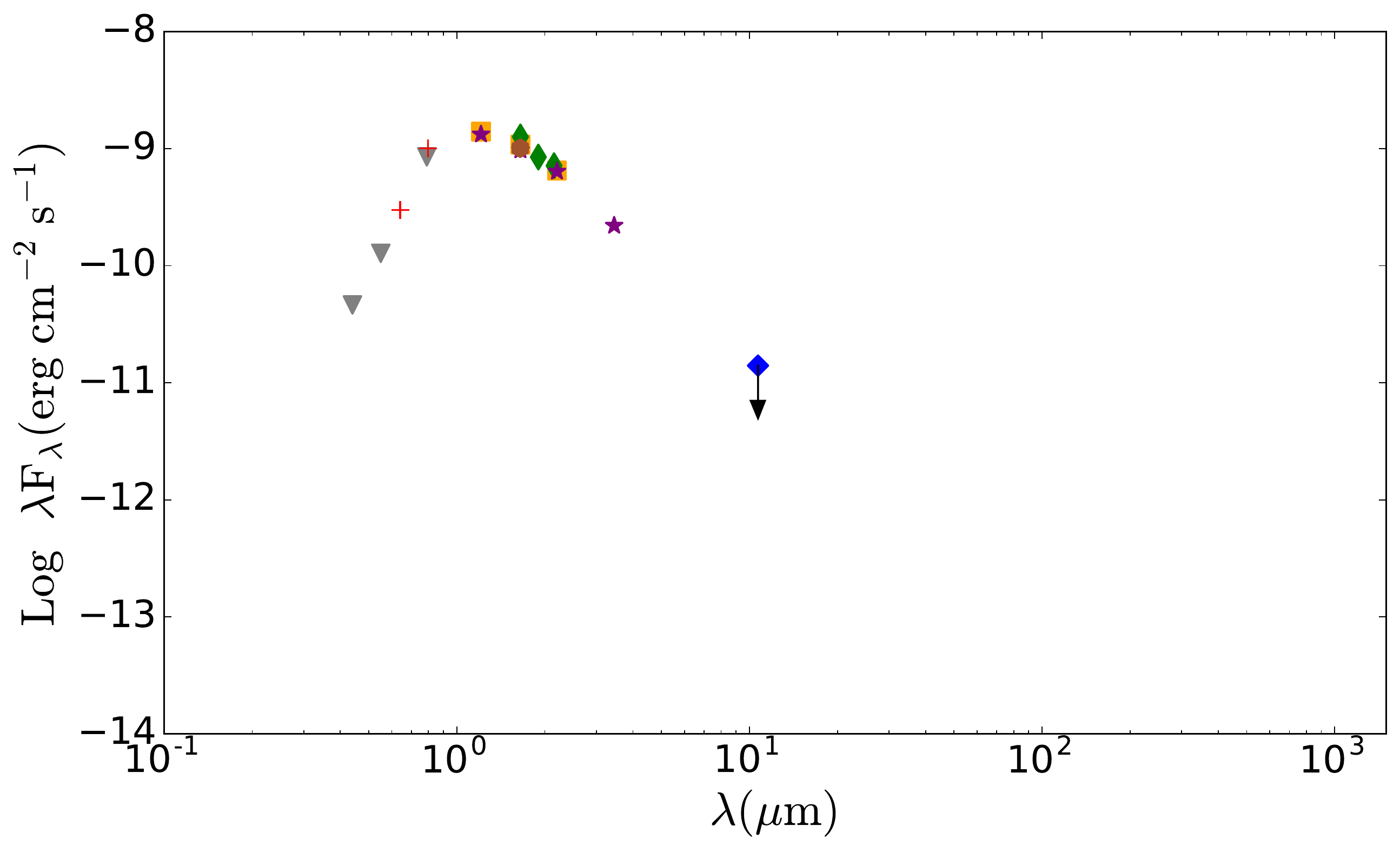}
\caption{SEDs based primarily on data from the literature; sources and references for the plotted points are given in bold face in Table 13.
Top plot shows TWA 3A and the bottom TWA 3B.}
\label{seds}
\end{figure}

\begin{figure}
\centering
\includegraphics[width=6in]{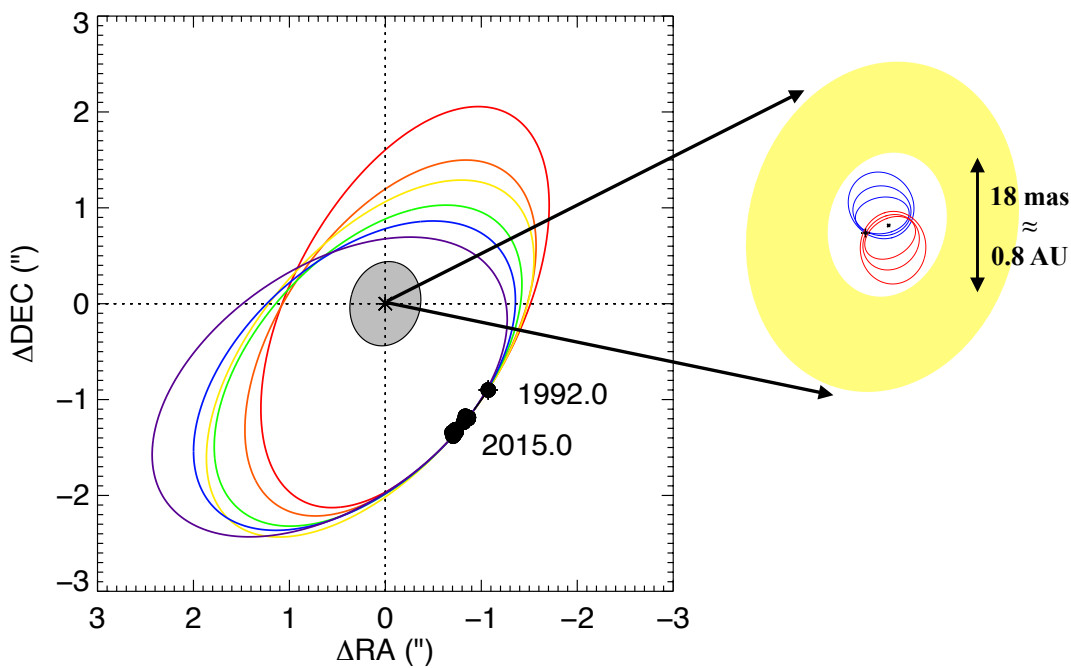}
\caption{Binary component orbits and disk
orientation for the TWA 3 system.  The gray ellipse shows the
orientation and outer radius of 20 AU for the 880 $\mu$m circumbinary dust disk around
TWA 3A, deconvolved from the beam size (Andrews et al. 2010).  The insert shows a blow up of the
inner cavity of tenuous gas with an inside radius of 0.4 AU and an
outer radius of 1.0 AU (Andrews et al. 2010).
To convert the physical dimensions used by Andrews et al. to an
angular scale for comparison with our orbit families, we used the
distance cited by Andrews et al., 45~pc.}
\label{phs}
\end{figure}

\clearpage

\begin{deluxetable}{lr}
\tablecaption{TWA 3 System Properties}
\tablewidth{0pt}
\tablehead{}
\startdata
RA (J2000.0)& 11h10m27.88s\\
Dec (J2000.0)& $-$37d31m52.0s\\
Distance (pc) & 34$\pm$4\tablenotemark{a}\\
A--B Separation (2015) & 1\farcs55\tablenotemark{b} \\
A--B Position angle (2015) & 207$^{\circ}$\tablenotemark{b} \\
$V$ (mag) & 12.04$\pm$0.01\tablenotemark{c} \\
$J$ (mag) & 7.651$\pm$0.019\tablenotemark{d}\\
$H$ (mag) & 7.041$\pm$0.027\tablenotemark{d}\\
$K_s$ (mag) & 6.774$\pm$0.020\tablenotemark{d}\\
A$_V$ & 0.01\tablenotemark{e}\\
%Aa SpT & M4\\
%Ab SpT & M4.5\\
%B SpTy & M3\\
\enddata

\tablenotetext{a}{Mamajek (2005)}
\tablenotetext{b}{Tokovinin et al. (2015)}
\tablenotetext{c}{Torres et al. (2000)}
\tablenotetext{d}{2MASS}
\tablenotetext{e}{McJunkin et al. (2014)}

\end{deluxetable}

\begin{deluxetable}{lr}
\tablecaption{Facilities and Observations}
\tablewidth{0pt}
\tablehead{}
\startdata
1.5 m Tillinghast $+$ echelle (CfA) & $R=35,000$ optical spectroscopy\\
4.5 m MMT $+$ echelle (CfA) & $R=35,000$ optical spectroscopy\\
1.5 m ESO $+$ FEROS & $R=44,000$ optical spectroscopy\\
2.5 m du Pont $+$ echelle & $R=40,000$ optical spectroscopy\\
10 m Keck II $+$ NIRSPEC & $R=30,000$ IR spectroscopy\\
10 m Keck II $+$ NIRSPAO & $R=30,000$ IR spectroscopy with AO\\
4.3 m DCT $+$ LMI & BVRI imaging\\
6.0 m Magellan $+$ Clio & HK AO imaging \\
\enddata 
\end{deluxetable}

\begin{deluxetable}{lll}
\tablecaption{Keck Infrared Spectroscopy}
\tablewidth{0pt}
\tablehead{
\colhead{UT} & \colhead{Component} & \colhead{  } \\
\colhead{Date} & \colhead{Observed} & \colhead{AO?}
}
\startdata
2002 Dec 22 & A, B &\\
2003 Feb 08 & A, B &\\
2003 Apr 20 & A & yes\\
2004 Jan 26 & A &\\
2004 Dec 26 & A &\\
2005 Feb 22 & A &\\
2009 Dec 06 & A, B & yes \\
2010 Feb 24 & A &\\
2010 Dec 12 & A, B & yes \\
2011 Feb 05 & A &\\
2011 Feb 09 & A &\\
\enddata

\end{deluxetable}

\begin{deluxetable}{lllllll} 
%\tabletypesize{\small}
\tabletypesize{\scriptsize}
\tablewidth{0pt}
\tablecaption{Position Measurements of TWA 3B relative to TWA 3A} 
\tablehead{
\colhead{UT Date} & \colhead{BY\tablenotemark{a}} & \colhead{$\rho$ (mas)} & \colhead{P.A.($\degr$)} & \colhead{Telescope} & \colhead{Filter} & \colhead{Flux Ratio}}
\startdata 
2014 Apr 21 02:10 & 2014.3030 & 1.5440 $\pm$ 0.0048  &  207.65 $\pm$ 0.18   & Magellan & $H$ &  0.689 $\pm$ 0.017  \\
                  &           &                      &                        &          & $K_s$ &  0.650 $\pm$ 0.034  \\
2014 Apr 08 05:45 & 2014.2678 & 1.5124 $\pm$ 0.0181  &  207.387 $\pm$ 0.687   & DCT      & $B$ &  0.8185 $\pm$ 0.0070  \\
                  &           &                      &                        &          & $V$ &  0.7543 $\pm$ 0.0030  \\
                  &           &                      &                        &          & $R$ &  0.7216 $\pm$ 0.0023  \\
                  &           &                      &                        &          & $I$ &  0.6987 $\pm$ 0.0049  \\
\enddata 

\tablenotetext{a}{Besselian Year.}

\label{tab.sepPA}
\end{deluxetable}

\clearpage

\begin{deluxetable}{lcrcc}
\tablewidth{0pc}
\tablecaption{Template Spectral Type Standard Stars\label{spts}}
\tablehead{
\colhead{$  $} & \colhead{Spectral} & \colhead{RV} & \colhead{$v \sin i$} &  \colhead{T$_{eff}$\tablenotemark{d}}\\
\colhead{Object} & \colhead{Type} & \colhead{(\kms)} & \colhead{(\kms)} &  \colhead{K}}
\startdata
Infrared & & &\\
\hline
GJ 15A & M3\tablenotemark{a}  & 9.6 & $<$2.5\tablenotemark{b} & 3453$\pm$86 \\
GJ 402 & M4 & $-$3.1 & $<$2.5\tablenotemark{b} & 3238$\pm$60 \\
GJ 669B & M4.5 & $-$36.8  & 6.1\tablenotemark{b} & 3131$\pm$85 \\
\hline
Optical & & &\\
\hline
GJ 48 & M3.0V & 2.49 & $<$2.5\tablenotemark{c} & 3453$\pm$86  \\
GJ 699 & M4.0V & $-$108.77 & $<$2.5\tablenotemark{c} & 3238$\pm$60
\enddata

\tablenotetext{a}{Prato (2007)}
\tablenotetext{b}{Reiners et al. (2012)}
\tablenotetext{c}{Browning et al. (2010)}
\tablenotetext{d}{From or derived from Mann et al. (2015)}

\end{deluxetable}

\begin{deluxetable}{lcrrr}
%\tabletypesize{\scriptsize}
\tablewidth{0pc}
\tablecaption{Keck IR Radial Velocity Measurements of TWA 3.\label{tab:keck_rvs}}
\tablehead{
\colhead{HJD} &
\colhead{Orbital} &
\colhead{$RV_{\rm Aa}$} &
\colhead{$RV_{\rm Ab}$} &
\colhead{$RV_{\rm B}$} \\
\colhead{(2,400,000$+$)} &
\colhead{phase} &
\colhead{(\kms)} &
\colhead{(\kms)} &
\colhead{(\kms)}}
\startdata
 52631.1403  &  0.895  &     31.22  &  $-$16.84  &  7.14     \\
 52679.0055  &  0.267  &   $-$3.66  &     21.03  &  7.09     \\
 52749.7738  &  0.296  &      0.00  &     22.07  &  \nodata  \\
 53031.0457  &  0.361  &      1.53  &     17.70  &  \nodata  \\
 53366.1060  &  0.967  &     32.25  &  $-$20.59  &  \nodata  \\
 53423.9356  &  0.625  &     12.45  &      3.01  &  \nodata  \\
 55172.1516  &  0.748  &     18.68  &   $-$4.97  &  7.64     \\
 55251.9446  &  0.036  &   $-$7.34  &     27.51  &  \nodata  \\
 55543.1431  &  0.385  &      1.81  &     15.55  &  7.79     \\
 55597.9402  &  0.956  &     34.40  &  $-$20.58  &  \nodata  \\
 55601.9331  &  0.071  &  $-$12.17  &     33.82  &  \nodata  \\
\enddata
\tablecomments{Radial velocity uncertainties are $\sigma_{\rm Aa} =
  0.63~\kms$, $\sigma_{\rm Ab} = 0.85~\kms$, and $\sigma_{\rm B} =
  0.59~\kms$ (see text).}
\end{deluxetable}

\begin{deluxetable}{lcrrrccc}
%\tabletypesize{\scriptsize}
\tablewidth{0pc}
\tablecaption{CfA Radial Velocity Measurements of TWA 3.\label{tab:cfa_rvs}}
\tablehead{
\colhead{HJD} &
\colhead{Orbital} &
\colhead{$RV_{\rm Aa}$} &
\colhead{$RV_{\rm Ab}$} &
\colhead{$RV_{\rm B}$} &
\colhead{$\sigma_{\rm Aa}$} &
\colhead{$\sigma_{\rm Ab}$} &
\colhead{$\sigma_{\rm B}$} \\
\colhead{(2,400,000$+$)} &
\colhead{phase} &
\colhead{(\kms)} &
\colhead{(\kms)} &
\colhead{(\kms)} &
\colhead{(\kms)} &
\colhead{(\kms)} &
\colhead{(\kms)}}
\startdata
 50828.9583  &   0.225  &   $-$4.28  &     26.71  &   5.94   &    4.10  &   6.66  &   3.95 \\
 51177.0639  &   0.205  &   $-$3.93  &     23.00  &   7.29   &    2.90  &   4.71  &   2.80 \\
 51237.9189  &   0.950  &     39.18  &  $-$13.93  &  12.90   &    2.65  &   4.30  &   2.55 \\
 51595.9255  &   0.215  &   $-$4.61  &     24.70  &  10.15   &    2.65  &   4.30  &   2.55 \\
 51619.8713  &   0.901  &     33.65  &  $-$14.36  &   9.47   &    2.77  &   4.49  &   2.67 \\
 51621.8603  &   0.958  &     37.45  &  $-$25.56  &   4.56   &    2.77  &   4.49  &   2.67 \\
 51682.7086  &   0.703  &     16.36  &      3.66  &  14.07   &    2.37  &   3.84  &   2.28 \\
 51917.0057  &   0.420  &      0.90  &     23.18  &  10.16   &    2.77  &   4.49  &   2.67 \\
 51945.9559  &   0.250  &   $-$3.11  &     28.13  &   9.82   &    2.29  &   3.72  &   2.21 \\
 51947.9030  &   0.306  &   $-$1.91  &     26.98  &   8.36   &    4.10  &   6.66  &   3.95 \\
 51971.8490  &   0.993  &     17.30  &      2.20  &   7.61   &    2.65  &   4.30  &   2.55 \\
 51972.8516  &   0.021  &      0.66  &     24.23  &  10.28   &    2.77  &   4.49  &   2.67 \\
 51973.8552  &   0.050  &   $-$9.99  &     36.49  &  11.10   &    3.06  &   4.97  &   2.95 \\
 52008.7574  &   0.051  &   $-$3.75  &     36.49  &   7.67   &    2.29  &   3.72  &   2.21 \\
 52009.7560  &   0.079  &  $-$12.87  &     34.92  &   8.09   &    2.29  &   3.72  &   2.21 \\
 52010.7549  &   0.108  &  $-$10.12  &     37.06  &   6.40   &    2.77  &   4.49  &   2.67 \\
 52011.7852  &   0.138  &   $-$9.87  &     32.41  &   7.45   &    2.16  &   3.51  &   2.08 \\
 52034.6899  &   0.794  &     23.48  &   $-$6.55  &   9.00   &    2.10  &   3.42  &   2.02 \\
 52036.6749  &   0.851  &     30.59  &   $-$3.31  &  12.04   &    2.23  &   3.61  &   2.15 \\
 52037.6722  &   0.880  &     32.22  &  $-$13.52  &   9.91   &    2.37  &   3.84  &   2.28 \\
 52038.6889  &   0.909  &     30.09  &  $-$17.56  &   3.97   &    2.65  &   4.30  &   2.55 \\
 52039.7160  &   0.938  &     34.12  &  $-$23.75  &   9.06   &    3.24  &   5.27  &   3.12 \\
 52336.8646  &   0.458  &      4.67  &      5.77  &  14.82   &    4.10  &   6.66  &   3.95 \\
 52360.7904  &   0.144  &   $-$4.51  &     33.18  &   8.67   &    2.55  &   4.13  &   2.46 \\
 52395.6929  &   0.145  &  $-$11.35  &     23.39  &   6.37   &    3.47  &   5.63  &   3.35 \\
 52653.9864  &   0.550  &      4.61  &      5.81  &  14.23   &    3.75  &   6.08  &   3.62 \\
 52722.7959  &   0.523  &      8.55  &     10.74  &  11.38   &    2.90  &   4.71  &   2.80 \\
 52771.6831  &   0.925  &     28.34  &  $-$20.69  &   5.19   &    3.24  &   5.27  &   3.12 \\
 53016.9925  &   0.958  &     36.04  &  $-$25.21  &   8.23   &    1.87  &   3.04  &   1.80 \\
 53035.9465  &   0.501  &      9.31  &     14.46  &   4.42   &    2.45  &   3.98  &   2.36 \\
 53127.7003  &   0.132  &   $-$5.98  &     34.88  &   9.67   &    2.16  &   3.51  &   2.08 \\
 53479.7427  &   0.225  &   $-$4.89  &     15.67  &   6.14   &    2.90  &   4.71  &   2.80 \\
 53781.9036  &   0.889  &     35.70  &  $-$16.91  &  10.66   &    2.37  &   3.84  &   2.28 \\
 53864.6736  &   0.262  &   $-$1.47  &     20.14  &   7.03   &    2.55  &   4.13  &   2.46 \\
 54136.9247  &   0.067  &  $-$10.99  &     31.18  &   9.40   &    2.77  &   4.49  &   2.67 \\
\enddata
\tablecomments{Radial velocity uncertainties for the TWA 3A components
  were determined iteratively from our combined orbital solution (see
  text), and account for the varying strength of each spectrum.}
\end{deluxetable}

\begin{deluxetable}{lcrrr}
%\tabletypesize{\scriptsize}
\tablewidth{0pc}
\tablecaption{FEROS radial velocity measurements of TWA 3.\label{tab:feros_rvs}}
\tablehead{
\colhead{HJD} &
\colhead{Orbital} &
\colhead{$RV_{\rm Aa}$} &
\colhead{$RV_{\rm Ab}$} &
\colhead{$RV_{\rm B}$} \\
\colhead{(2,400,000$+$)} &
\colhead{phase} &
\colhead{(\kms)} &
\colhead{(\kms)} &
\colhead{(\kms)}}
\startdata
 51260.5221  &  0.598  &    13.23  &      2.40  &  10.11 \\
 51331.5915  &  0.636  &    10.88  &   $-$0.99  &  15.45 \\
 51621.5378  &  0.949  &    38.05  &  $-$17.73  &  10.11 \\
 51623.6000  &  0.008  &    15.76  &      9.47  &   6.49 \\
 51624.5913  &  0.037  &  $-$4.49  &     32.46  &  12.74 \\
 51625.5186  &  0.063  &  $-$7.27  &     37.20  &  11.83 \\
 51733.4665  &  0.158  &  $-$4.76  &     32.97  &  12.54 \\
 51737.5133  &  0.274  &  $-$0.56  &     28.24  &  11.77 \\
\enddata
\tablecomments{Radial velocity uncertainties are $\sigma_{\rm Aa} =
  2.61~\kms$, $\sigma_{\rm Ab} = 3.59~\kms$, and $\sigma_{\rm B} =
  2.60~\kms$ (see text).}
\end{deluxetable}

\begin{deluxetable}{lcrrr}
%\tabletypesize{\scriptsize}
\tablewidth{0pc}
\tablecaption{du Pont Radial Velocity Measurements of TWA 3.\label{tab:dupont_rvs}}
\tablehead{
\colhead{HJD} &
\colhead{Orbital} &
\colhead{$RV_{\rm Aa}$} &
\colhead{$RV_{\rm Ab}$} &
\colhead{$RV_{\rm B}$} \\
\colhead{(2,400,000$+$)} &
\colhead{phase} &
\colhead{(\kms)} &
\colhead{(\kms)} &
\colhead{(\kms)}}
\startdata
 53765.6694  &  0.422    &      4.97  &     11.56  &     16.68 \\
 53765.8423\tablenotemark{a}  & \nodata    &  \nodata    &  \nodata    &     10.56 \\
 53765.8556\tablenotemark{b}  &  0.427    &      6.52  &     15.92  &  \nodata   \\
 53766.7060  &  0.452    &      5.46  &     13.08  &     13.79 \\
 53767.7278  &  0.481    &      6.77  &     14.30  &     18.40 \\
 53767.8467  &  0.485    &      7.50  &     13.55  &     16.57 \\
 53769.7425  &  0.539    &     13.42  &      6.66  &      8.52 \\
 53770.7760  &  0.569    &     13.38  &      4.58  &     15.00 \\
 53771.7166  &  0.595    &     13.40  &      5.98  &      8.05 \\
\enddata
\tablecomments{Radial velocity uncertainties are $\sigma_{\rm Aa} =
  1.46~\kms$, $\sigma_{\rm Ab} = 2.34~\kms$, and $\sigma_{\rm B} =
  3.95~\kms$ (see text).}
\tablenotetext{a}{Spectrum of TWA 3B alone.}
\tablenotetext{b}{Spectrum of TWA 3A alone.}
\end{deluxetable}

\begin{deluxetable}{lccc}
\tablewidth{0pc}
\tablecaption{Spectroscopic Orbital Solutions for TWA 3A\label{tab:orbit}}
\tabletypesize{\small}
\tablehead{
\colhead{\hfil~~~~~~~~~~~~~~~~~Parameter~~~~~~~~~~~~~~~~~~} & 
\colhead{Keck} &
\colhead{CfA} &
\colhead{Combined\tablenotemark{a}}
}
\startdata
$P$ (days)\dotfill                                   &   34.8799~$\pm$~0.0022\phn     &   34.8742~$\pm$~0.0088\phn     &   34.87846~$\pm$~0.00090\phn               \\
$\gamma$ (\kms)\dotfill                              &   $+$8.63~$\pm$~0.18\phs       &   $+$9.61~$\pm$~0.41\phs       &   $+$10.17~$\pm$~0.40\tablenotemark{b}\phs \\
$K_{\rm Aa}$ (\kms)\dotfill                              &     23.38~$\pm$~0.36\phn       &     23.33~$\pm$~0.68\phn       &      23.28~$\pm$~0.26\phn                  \\
$K_{\rm Ab}$ (\kms)\dotfill                              &     27.74~$\pm$~0.49\phn       &      27.4~$\pm$~1.1\phn        &      27.68~$\pm$~0.36\phn                  \\
$e$\dotfill                                          &    0.6323~$\pm$~0.0091         &     0.636~$\pm$~0.019          &     0.6280~$\pm$~0.0060                    \\
$\omega_{Aa}$ (deg)\dotfill                          &      81.2~$\pm$~1.5\phn        &      76.5~$\pm$~2.6\phn        &       80.5~$\pm$~1.2\phn                   \\
$T$ (HJD$-$2,400,000)\tablenotemark{c}\dotfill       &  52704.53~$\pm$~0.13\phm{2222} &  52704.24~$\pm$~0.20\phm{2222} &  52704.554~$\pm$~0.063\phm{2222}           \\
$\Delta RV$(CfA$-$Keck) (\kms)\dotfill               &         \nodata                &          \nodata               &    $+$1.53~$\pm$~0.43\phs              \\
$\Delta RV$(CfA$-$FEROS) (\kms)\dotfill              &         \nodata                &          \nodata               &    $-$1.15~$\pm$~0.87\phs                   \\
$\Delta RV$(CfA$-$du Pont) (\kms)\dotfill             &         \nodata                &          \nodata               &    $+$0.15~$\pm$~0.61\phs                  \\[-0.5ex]
%\tablevspace{8pt}
\cutinhead{Derived quantities}
$M_{\rm Aa}\sin^3 i$ ($M_{\sun}$)\dotfill                &    0.1218~$\pm$~0.0055         &     0.117~$\pm$~0.010          &     0.1224~$\pm$~0.0042                    \\
$M_{\rm Ab}\sin^3 i$ ($M_{\sun}$)\dotfill                &    0.1027~$\pm$~0.0043         &    0.0999~$\pm$~0.0071         &     0.1030~$\pm$~0.0032                    \\
$q\equiv M_{\rm Ab}/M_{\rm Aa}$\dotfill                      &      0.843~$\pm$~0.018         &     0.851~$\pm$~0.040          &      0.841~$\pm$~0.014                     \\
$a_{\rm Aa}\sin i$ ($10^6$~km)\dotfill                   &      8.69~$\pm$~0.14           &      8.64~$\pm$~0.24           &       8.69~$\pm$~0.11                      \\
$a_{\rm Ab}\sin i$ ($10^6$~km)\dotfill                   &     10.31~$\pm$~0.19\phn       &     10.15~$\pm$~0.40\phn       &      10.33~$\pm$~0.15\phn                  \\
$a \sin i$ ($R_{\sun}$)\dotfill                      &     27.31~$\pm$~0.39\phn       &     27.01~$\pm$~0.69\phn       &      27.34~$\pm$~0.29\phn                  \\[-0.5ex]
%\tablevspace{8pt}
\cutinhead{Other quantities pertaining to the fit}
$N_{\rm Aa}$~,~$N_{\rm Ab}$, Keck\dotfill                    &          11~,~11               &          \nodata               &           11~,~11                          \\
$N_{\rm Aa}$~,~$N_{\rm Ab}$, CfA\dotfill                     &         \nodata                &          35~,~35               &           35~,~35                          \\
$N_{\rm Aa}$~,~$N_{\rm Ab}$, FEROS\dotfill                   &         \nodata                &          \nodata               &            8~,~8                           \\
$N_{\rm Aa}$~,~$N_{\rm Ab}$, du Pont\dotfill                  &         \nodata                &          \nodata               &            8~,~8                           \\
Time span (days)\dotfill                             &          2970.8                &           3308.0               &           4773.0                           \\
$\sigma_{\rm Aa}$~,~$\sigma_{\rm Ab}$, Keck (\kms)\dotfill   &        0.73~,~1.05             &          \nodata               &         0.66~,~0.89                        \\
$\sigma_{\rm Aa}$~,~$\sigma_{\rm Ab}$, CfA (\kms)\dotfill    &         \nodata                &        2.74~,~4.64             &         2.74~,~4.45                        \\
$\sigma_{\rm Aa}$~,~$\sigma_{\rm Ab}$, FEROS (\kms)\dotfill  &         \nodata                &          \nodata               &         2.72~,~3.74                        \\
$\sigma_{\rm Aa}$~,~$\sigma_{\rm Ab}$, du Pont (\kms)\dotfill &         \nodata                &          \nodata               &         1.52~,~2.45                        \\
\enddata
\tablecomments{$\omega_{Ab}=\omega_{Aa}+180 = 260.5$ degrees.}
\tablenotetext{a}{Simultaneous solution using the Keck, CfA, FEROS, and du Pont data sets.}
\tablenotetext{b}{On the reference system of CfA.}
\tablenotetext{c}{Time of periastron passage nearest to the average of
all times of observation from the four data sets.}
\end{deluxetable}

%\clearpage

\begin{deluxetable}{lcc}
\tablewidth{0pc}
\tablecaption{Range of Orbital Parameters obtained for Wide
A--B Orbit Search\label{min_max}}
\tablehead{
\colhead{Parameter} & \colhead{Minimum} & \colhead{Maximum} }
\startdata
%\hline
$P$ (years) &    236 &   800 \\
$T$ (BY) &  1610 &   2386\\
$e$ &    0.00 &   0.80 \\
$a$ (\arcsec)  &      1.17  &    3.18\\
$i$ (degrees) &    108 &   172 \\
$\Omega$   (degrees) &     0 &   140 \\
\nodata &   320 & 360 \\
$\omega_B$ (degrees) &      0 &   360\\
$M$ ($M_{\odot}$)  &    0.32   &   2.00\\
\enddata
\tablecomments{While searching for orbital solutions, we added a
constraint that the total mass must be smaller than 2 $M_{\odot}$,
assuming a distance of 34 pc.  We placed an arbitrary upper limit of 800 yr on the period.}
\label{min_max}
\end{deluxetable}
\clearpage

\begin{deluxetable}{llllllll}
\tablewidth{0pc}
\tablecaption{Parameters for Wide A--B Orbits Plotted in Figure
5\label{AB_family}}
\tablehead{
\colhead{$P$} & \colhead{$T$} & \colhead{$e$} & \colhead{$a$} &
\colhead{$i$} & \colhead{$\Omega$} & \colhead{$\omega_B$} &
\colhead{$M_{tot}$}\\
\colhead{(years)} & \colhead{(BY)} & \colhead{$  $} & \colhead{($''$)}
& \colhead{(degrees)} &
\colhead{ (degrees)} & \colhead{(degrees)} & \colhead{($M_{\odot}$)}}
\startdata
785.76  & 2321.3  & 0.1694  & 2.321  & 120.67  & 331.95  & 260.46  & 0.796\\
699.96  & 1707.0  & 0.2490  & 2.160  & 122.30  & 324.51  & 298.97  & 0.808\\
766.13  & 1761.2  & 0.3215  & 2.291  & 122.11  & 137.13  & 135.74  & 0.805\\
671.73  & 1805.0  & 0.3976  & 2.100  & 124.91  & 133.24 &  133.43  & 0.807\\
685.21  & 1835.6  & 0.4671  & 2.119  & 126.52  & 126.39  & 135.42  & 0.797\\
748.55  & 1868.4  & 0.5602  & 2.245  & 128.22  & 114.51  & 135.00  & 0.793\\
\enddata
\tablecomments{The total masses were computed assuming a distance of
34 pc from Mamajek (2005).}
\end{deluxetable}

\clearpage

\begin{deluxetable}{cccccccc}
\tablewidth{0pt} 
\tablecaption{TWA 3 Magnitudes and Flux Densities \label{tbl-13}}
%\rotate
\tabletypesize{\scriptsize}
\tablehead{
\colhead{$ $} & \colhead{$\lambda$} & \colhead{A} & \colhead{B} & \colhead{A$+$B} & \colhead{$\lambda$F$_{\lambda}$A} & \colhead{$\lambda$F$_{\lambda}$B} & \colhead{$ $}\\
\colhead{Filter} & \colhead{$(\mu \mathrm{m})$} & \colhead{(mag)} & \colhead{(mag)} & \colhead{(mag)} & \colhead{(erg/cm$^2$/s)} & \colhead{(erg/cm$^2$/s)} & \colhead{Source}}
\startdata
U & 0.3656 & \nodata & \nodata & 14.05 & \nodata & \nodata & 1 \\
U & 0.3656 & \nodata & \nodata & 14.21 & \nodata & \nodata & 2 \\
U & 0.3656 & \nodata & \nodata & 14.27 & \nodata & \nodata & 3 \\
B & 0.4353 & \nodata & \nodata & 13.537$\pm$0.02 & \nodata & \nodata & 4 \\
B & 0.4353 & \nodata & \nodata & 13.58 & \nodata & \nodata & 1 \\
B & 0.4353 & \nodata & \nodata & 13.52 & \nodata & \nodata & 2 \\
B & 0.4353 & \nodata & \nodata & 13.53 & \nodata & \nodata & 3 \\
\textbf{B} & \textbf{0.44} & \textbf{14.04} & \textbf{14.59} & \nodata & {\bf $7.669\times10^{-11}$ } & \textbf{$4.621\times10^{-11}$} & \textbf{5}\\
g' & 0.4639 & \nodata & \nodata & 12.74$\pm$0.02 & \nodata & \nodata & 4 \\
V & 0.5477 & \nodata & \nodata & 12.05$\pm$0.01 & \nodata & \nodata & 4 \\
V & 0.5477 & \nodata & \nodata & 12.06 & \nodata & \nodata & 1 \\
V & 0.5477 & \nodata & \nodata & 12.04 & \nodata & \nodata & 2 \\
V & 0.5477 & \nodata & \nodata & 12.04 & \nodata & \nodata & 3 \\
\textbf{V} & \textbf{0.55} & \textbf{12.57} & \textbf{13.07} & \nodata & \textbf{$2.022\times10^{-10}$} & \textbf{$1.276\times10^{-10}$} & \textbf{5} \\
r' & 0.6122 & \nodata & \nodata & 11.442$\pm$0.02 & \nodata & \nodata & 4 \\
Rc & 0.6407 & \nodata & \nodata & 10.72 & \nodata & \nodata & 1 \\
Rc & 0.6407 & \nodata & \nodata & 10.66 & \nodata & \nodata & 2 \\
Rc & 0.6407 & \nodata & \nodata & 10.69 & \nodata & \nodata & 3 \\
\textbf{Rc\tablenotemark{a}} & \textbf{0.6407} & \textbf{11.28$\pm$0.01} & \textbf{11.63$\pm$0.01} & \nodata & \textbf{$4.141\times10^{-10}$} & \textbf{$3.000\times10^{-10}$} & \textbf{1,2,3,6} \\
i' & 0.7439 & \nodata & \nodata & 9.826$\pm$0.05 & \nodata & \nodata & 4 \\
%i' & 0.7439 & \nodata & \nodata & \nodata & \nodata & \nodata & \nodata & 0.711 & \nodata & 0.37 & 0.01 & 26 \\
I & 0.79 & \nodata & \nodata & 9.29$\pm$0.03 & \nodata & \nodata & 7 \\
\textbf{I} & \textbf{0.79} & \textbf{9.72} & \textbf{10.0} & \nodata & \textbf{$1.101\times10^{-09}$}  & \textbf{$8.510\times10^{-10}$} & \textbf{5} \\
Ic & 0.798 & \nodata & \nodata & 9.14 & \nodata & \nodata & 1 \\
Ic & 0.798 & \nodata & \nodata & 9.11 & \nodata & \nodata & 2 \\
Ic & 0.798 & \nodata & \nodata & 9.1 & \nodata & \nodata & 3 \\
\textbf{Ic\tablenotemark{a}} & \textbf{0.798} & \textbf{9.69$\pm$0.01} & \textbf{10.08$\pm$0.01} & \nodata & \textbf{$1.121\times10^{-09}$}  & \textbf{$1.004\times10^{-09}$} & \textbf{1,2,3,6} \\
%zÕ & 0.8896 & \nodata & \nodata & \nodata & \nodata & \nodata & \nodata & 0.685 & \nodata & 0.41 & 0.01 & 26 \\
\textbf{J} & \textbf{1.21} & \textbf{8.22$\pm$0.10} & \textbf{8.63$\pm$0.10} & \nodata & \textbf{$2.046\times10^{-09}$}  & \textbf{$1.403\times10^{-09}$}  & \textbf{8} \\
\textbf{J} & \textbf{1.21} & \textbf{8.25$\pm$0.03} & \textbf{8.69$\pm$0.03} & \nodata & \textbf{$1.991\times10^{-09}$}  & \textbf{$1.327\times10^{-09}$}  & \textbf{9} \\
J & 1.21 & \nodata & \nodata & 7.61 & \nodata & \nodata & 10 \\
J & 1.235 & \nodata & \nodata & 7.651$\pm$0.019 & \nodata & \nodata & 11 \\
J & 1.24 & \nodata & \nodata & 7.63$\pm$0.05 & \nodata & \nodata & 7 \\
%FeII & 1.644 & \nodata & \nodata & \nodata & \nodata & \nodata & \nodata & 0.69 & \nodata & 0.4 & \nodata & 17 \\
\textbf{F164N} & \textbf{1.65} & \textbf{7.52$\pm$0.01} & \textbf{7.9$\pm$0.01} & \nodata & \textbf{$1.790\times10^{-09}$}  & \textbf{$1.2623\times10^{-09}$}  & \textbf{12} \\
\textbf{H} & \textbf{1.648} & \textbf{7.53$\pm$0.05} & \textbf{8.15$\pm$0.07} & \nodata & \textbf{$1.788\times10^{-09}$}  & \textbf{$1.010\times10^{-09}$}  & \textbf{13} \\
\textbf{H} & \textbf{1.65} & \textbf{7.6$\pm$0.1} & \textbf{8.1$\pm$0.1} & \nodata & \textbf{$1.675\times10^{-09}$} & \textbf{$1.086\times10^{-09}$}  & \textbf{8} \\
\textbf{H} & \textbf{1.65} & \textbf{7.79$\pm$0.03} & \textbf{8.19$\pm$0.03} & \nodata & \textbf{$1.406\times10^{-09}$}  & \textbf{$9.727\times10^{-10}$} & \textbf{9} \\
H & 1.662 & \nodata & \nodata & 7.041$\pm$0.027 & \nodata & \nodata & 11 \\
\textbf{F190N} & \textbf{1.9} & \textbf{7.60$\pm$0.01} & \textbf{7.98$\pm$0.01} & \nodata & \textbf{$1.204\times10^{-09}$}  & \textbf{$8.488\times10^{-10}$}  & \textbf{12} \\
\textbf{F215N} & \textbf{2.15} & \textbf{7.37$\pm$0.01} & \textbf{7.79$\pm$0.01} & \nodata & \textbf{$1.055\times10^{-09}$} & \textbf{$7.163\times10^{-10}$}  & \textbf{12} \\
K & 2.159 & \nodata & \nodata & 6.77$\pm$0.02 & \nodata & \nodata & 11 \\
K & 2.16 & \nodata & \nodata & 6.77$\pm$0.08 & \nodata & \nodata & 7 \\
\textbf{K} & \textbf{2.2} & \textbf{7.28$\pm$0.07} & \textbf{7.80$\pm$0.07} & \nodata &\textbf{ $1.052\times10^{-09}$} & \textbf{$6.517\times10^{-10}$}  & \textbf{8} \\
%K & 2.2 & \nodata & \nodata & \nodata & \nodata & \nodata & \nodata & 0.625 & \nodata & 0.51 & \nodata & 5 \\
K & 2.2 & \nodata & \nodata & 6.73 & \nodata & \nodata & 10 \\
\textbf{K} & \textbf{2.2} & \textbf{7.39$\pm$0.03} & \textbf{7.82$\pm$0.03} & \nodata & \textbf{$9.507\times10^{-10}$}  & \textbf{$6.398\times10^{-10}$}  & \textbf{9} \\
%BrktGma & 2.166 & \nodata & \nodata & \nodata & \nodata & \nodata & \nodata & 0.68 & \nodata & 0.42 & \nodata & 17 \\
W1 & 3.35 & \nodata & \nodata & 6.601$\pm$0.038 & \nodata & \nodata & 14 \\
\textbf{L} & \textbf{3.45} & \textbf{7.05$\pm$0.08} & \textbf{7.6$\pm$0.08} & \nodata & \textbf{$3.659\times10^{-10}$}  & \textbf{$2.205\times10^{-10}$} & \textbf{9} \\
Spitzer & 3.6 & \nodata & \nodata & 6.49$\pm$0.02 & \nodata & \nodata & 15 \\
Spitzer & 4.5 & \nodata & \nodata & 6.37$\pm$0.02 & \nodata & \nodata & 15 \\
W2 & 4.6 & \nodata & \nodata & 6.341$\pm$0.021 & \nodata & \nodata & 14 \\
%M & 4.8 & \nodata & \nodata & \nodata & \nodata & \nodata & \nodata & 0.526 & \nodata & \nodata & \nodata & 5 \\
\textbf{M} & \textbf{4.8} & \textbf{6.8$\pm$0.1} & \nodata & \nodata & \textbf{$1.822\times10^{-10}$} & \nodata & \textbf{9} \\
Spitzer & 5.8 & \nodata & \nodata & 6.15$\pm$0.03 & \nodata & \nodata & 15 \\
Spitzer & 8.0 & \nodata & \nodata & 5.15$\pm$0.03 & \nodata & \nodata & 15 \\
\textbf{N} & \textbf{10} & \nodata &  \nodata & \nodata &  \textbf{$2.012\times10^{-10}$} &  \nodata & \textbf{16} \\
\textbf{N} & \textbf{10.7} & \nodata & \nodata & \nodata & \textbf{$2.523\times10^{-10}$}  & \textbf{$<1.402\times10^{-11}$}  & \textbf{17} \\
W3 & 11.6 & \nodata & \nodata & 3.876$\pm$0.016 & \nodata & \nodata & 14 \\
\textbf{IRAS} & \textbf{12} & \nodata &  \nodata & \nodata &  \textbf{$2.471\times10^{-10}$ \tablenotemark{b}}  & \nodata & \textbf{18} \\
\textbf{Q} & \textbf{18.2} & \nodata &  \nodata & \nodata &  \textbf{$8.242\times10^{-11}$}  &  \nodata & \textbf{17} \\
W4 & 22.1 & \nodata & \nodata & 1.734$\pm$0.014 & \nodata & \nodata & 14 \\
Spitzer & 24.0 & \nodata & \nodata & 1.62$\pm$0.04 & \nodata & \nodata & 15 \\
\textbf{IRAS} & \textbf{25} & \nodata  &  \nodata & \nodata & \textbf{$2.096\times10^{-10}$ \tablenotemark{b}} & \nodata & \textbf{18} \\
\textbf{IRAS} & \textbf{60} & \nodata  &  \nodata & \nodata &   \textbf{$4.597\times10^{-11}$ \tablenotemark{b}}  & \nodata & \textbf{18} \\
\textbf{IRAS} & \textbf{100} & \nodata  &  \nodata & \nodata &  \textbf{$3.543\times10^{-11}$ \tablenotemark{b}}    & \nodata & \textbf{18} \\
\textbf{Far-IR} & \textbf{100} & \nodata  &  \nodata & \nodata &  \textbf{$5.600\times10^{-12}$ \tablenotemark{b}}   & \nodata & \textbf{19} \\
\textbf{Far-IR} & \textbf{160} &  \nodata  &  \nodata & \nodata &  \textbf{$2.869\times10^{-12}$ \tablenotemark{b}}   & \nodata & \textbf{19} \\
\textbf{Sub-mm} & \textbf{880} &   \nodata  &  \nodata & \nodata &  \textbf{$8.523\times10^{-14}$ \tablenotemark{b}}    & \nodata & \textbf{20} \\
\enddata

\tablenotetext{a}{Rc and Ic magnitudes in bold face plotted in Figure 7 were determined from the average of the
Rc and Ic magnitudes from the literature in conjunction with the flux ratios for the R and I filters given in Table 4.}
\tablenotetext{b}{Unresolved flux densities for A$+$B; the majority of flux, an order of magnitude or more, is from the A component
and so is attributed to A in the SED plots, upper panel of Figure 7.}
\tablerefs{(1) De la Reza et al. (1989); (2) Gregorio-Hetem et al. (1992); (3) Torres et al. (2000); (4) Munari et al. (2014);
(5) Torres et al. (2006); (6) This Work; (7) Leggett et al. (2001); (8) Webb et al. (1999); (9) Geoffray \& Monin (2001);
(10) Zuckerman et al. (2001); (11) Cutri et al. (2003); (12) Weintraub et al. (2000); (13) Brandeker et al. (2003);
(14) Cutri et al. (2012); (15) Luhman et al. (2010); (16) Metchev et al. (2004); (17) Jayawardhana et al. (1999); 
(18) Beichman et al. (1988); (19) Riviere-Marichalar et al. (2013); (20) Andrews et al. (2010).}

\end{deluxetable}

\end{document}